\documentclass[nofootinbib,aps,floatfix,superscriptaddress]{revtex4}

\usepackage[normalem]{ulem}
\usepackage{amssymb,amsmath,epsfig,slashed}
\usepackage[usenames,dvipsnames]{color}

\usepackage[colorlinks=true, linkcolor=blue, citecolor=blue, urlcolor=blue]{hyperref}

\begin{document}

%%%%%%%%%%%%%%%%%%%%%%%%%%%%%%%%%%%%%%%%%%%%%%%%%%%%%
\title{Gluon Generalized TMD signatures at the EIC from exclusive heavy\\ (axial-)vector meson production}
%%%%%%%%%%%%%%%%%%%%%%%%%%%%%%%%%%%%%%%%%%%%%%%%%%%%%

\author{Shohini Bhattacharya}
\email{shohinib@uconn.edu}
\affiliation{Department of Physics, University of Connecticut, Storrs, CT 06269, USA}

\author{David DeAngelo}
\email{david.deangelo@uconn.edu}
\affiliation{Department of Physics, University of Connecticut, Storrs, CT 06269, USA}

\author{Lei Yang}
\email{lei.yang@mail.sdu.edu.cn}
\affiliation{School of Physics and Key Laboratory of Particle Physics and Particle Irradiation (MOE), Shandong University, QingDao, Shandong, 266237, China}

\author{Duxin Zheng}
\email{duxin.zheng@iat.cn}
\affiliation{Shandong Institute of Advanced Technology, Jinan, Shandong, 250100, China}

\author{Jian Zhou}
\email{jzhou@sdu.edu.cn}
\affiliation{School of Physics and Key Laboratory of Particle Physics and Particle Irradiation (MOE), Shandong University, QingDao, Shandong, 266237, China}
\affiliation{Southern Center for Nuclear-Science Theory (SCNT), Institute of Modern Physics, Chinese Academy of Sciences, HuiZhou, Guangdong
516000, China\vspace{0.2cm}}

%%%%%%%%%%%%%%%%%%%%%%%%%%%%%%%%%%%%%%%%%%%%%%%%%%%%%    
\begin{abstract}
Potential experimental signatures of gluon generalized transverse momentum-dependent distributions (GTMDs) are proposed via exclusive heavy (axial-)vector meson production in lepton–proton collisions. Within the framework of collinear twist-3 factorization, we show that specific azimuthal-angle-dependent observables can provide sensitivity to the gluon GTMDs $F_{1,4}^g$ and $G_{1,1}^g$, which are related to partonic orbital angular momentum and spin–orbit correlations, respectively. These functions represent a unique sector of nucleon structure with no counterparts in the generalized parton distribution or transverse-momentum-dependent frameworks. We show that interference between different virtual-photon polarizations leads to distinct azimuthal modulations, including the polarization-independent $\cos 2\phi$ and polarization-dependent $\sin 2\phi$ terms, with $\phi$ defined as the angle between the lepton scattering plane and the hadron production plane. These observables provide signatures of the elusive gluon GTMDs $F_{1,4}^g$ and $G_{1,1}^g$, opening a new channel to access the spin structure of the nucleon at the future Electron–Ion Collider.
\end{abstract}
%%%%%%%%%%%%%%%%%%%%%%%%%%%%%%%%%%%%%%%%%%%%%%%%%%%%%

\date{\today}

\maketitle

%%%%%%%%%%%%%%%%%%%%%%%%%%%%%%%%%%%%%%%%%%%%%%%%%%%%%   
\section{Introduction} 
\label{s:intro}
%%%%%%%%%%%%%%%%%%%%%%%%%%%%%%%%%%%%%%%%%%%%%%%%%%%%%   
The quest to map the multi-dimensional structure of the nucleon has evolved from the initial discovery of the ``spin crisis'' to a sophisticated program aimed at understanding the transverse motion and spatial distribution of partons. While the longitudinal spin structure, encoded in quark and gluon helicity distributions, has been extensively studied through polarized deep inelastic scattering (DIS) and proton-proton collision data~\cite{STAR:2014wox, deFlorian:2014yva, Nocera:2014gqa, STAR:2021mqa}, the orbital angular momentum (OAM) remains one of the least constrained components of the Jaffe-Manohar spin sum rule~\cite{Jaffe:1989jz}. Accessing OAM is challenging because it inherently involves correlations between partonic transverse momentum and spatial position (impact parameter). Capturing such correlations requires going beyond traditional collinear or transverse-momentum-dependent (TMD) frameworks to the more general Wigner distribution formalism~\cite{Lorce:2011kd, Lorce:2011ni}.

Within this context, Generalized Transverse Momentum-Dependent distributions (GTMDs)~\cite{Meissner:2009ww} serve as the ``mother distributions'' of the nucleon, providing a unified mapping that reduces to Generalized Parton Distributions (GPDs) and TMDs in appropriate kinematic limits. A pivotal discovery was the direct correspondence between the GTMD $F_{1,4}$ and the partonic orbital angular momentum (OAM)~\cite{Lorce:2011kd, Hatta:2011ku}. Subsequent investigations have extended this reach to spin-orbit correlations, parameterized by functions such as $G_{1,1}$~\cite{Lorce:2014mxa,Bhattacharya:2024sno}, which, in the small-$x$ regime, have been shown to encode a significant anti-alignment between partonic helicity and OAM~\cite{Bhattacharya:2022vvo}. Importantly, these functions represent a unique domain of nucleon structure; unlike GPDs or TMDs, they capture the non-trivial correlations between partonic transverse momentum and the nucleon's recoil, and consequently vanish upon integration over either transverse degree of freedom.

Experimental access to these elusive GTMDs requires processes sensitive to both the parton’s transverse momentum and the momentum transfer to the nucleon. Over the past decade, several measurement strategies have emerged. Initial theoretical interest focused on diffractive dijet production in $ep$ and $eA$ collisions~\cite{Ji:2016jgn, Hatta:2016aoc}, which primarily probes gluon GTMDs in the small-$x$ regime but faces significant challenges from soft gluon bremsstrahlung backgrounds. In 2017, the exclusive double Drell-Yan process, $\pi N\rightarrow(\ell_{1}^{-}\ell_{1}^{+})(\ell_{2}^{-}\ell_{2}^{+})N^{\prime}$, was identified as the first physical channel providing access to quark GTMDs across the entire $x$ range~\cite{Bhattacharya:2017bvs}. However, the cross section for this process is severely suppressed by the electromagnetic coupling associated with two di-lepton pairs, rendering experimental extraction extremely challenging. More recently, exclusive $\pi^0$ production~\cite{Bhattacharya:2023yvo,Bhattacharya:2023hbq} and refined dijet analyses~\cite{Bhattacharya:2022vvo,Bhattacharya:2024sck,Kovchegov:2024wjs} have been proposed to target specific sectors of the Wigner distributions. These limitations—whether due to low rates or restricted kinematic sensitivity—underscore the need for high-rate observables capable of directly isolating gluon GTMDs.

In this work, we demonstrate that exclusive vector ($V$) and axial-vector ($AV$) meson production in electron--proton collisions,
%%%%%%%%%%%%%%%%%%%%%%
\begin{equation}
e + p \to e' + p' + m 
\quad \big( m = V/AV, \; V = J/\psi, \Upsilon, \dots, \; AV = \chi_{c1}, \chi_{b1}, \dots \big),
\end{equation}
%%%%%%%%%%%%%%%%%%%%%%
provides a robust, high-rate channel for isolating the gluon GTMDs $F^g_{1,4}$ and $G^g_{1,1}$. For clarity, in the main body of the paper we present results for vector mesons, while the more complex axial-vector results are provided in a dedicated appendix.  
Within the framework of collinear factorization, we find that, for vector mesons, interference between different virtual-photon polarizations leads to distinct azimuthal modulations, including the polarization-independent $\cos 2\phi$ and polarization-dependent $\sin 2\phi$ terms, with $\phi$ defined as the angle between the lepton scattering plane and the hadron production plane. These modulations provide clear signatures of the gluon GTMDs $F^g_{1,4}$ and $G^g_{1,1}$.  
Our calculation focuses on the kinematic twist-3 contributions arising from the $k_\perp$ and $\Delta_\perp$ expansions of the hard kernel, where $k_\perp$ denotes the intrinsic transverse momentum of the parton inside the proton, and $\Delta_\perp$ represents the transverse momentum transfer to the recoiling proton.
The azimuthal modulations we propose result entirely from interference between these twist-3 contributions.  
Our results provide a potential roadmap for the upcoming Electron–Ion Collider (EIC), where high-luminosity measurements will enable a systematic study of canonical orbital angular momentum and spin–orbit correlations. Further extensions of this work, including genuine twist-3 effects from three-gluon correlation functions, a more complete treatment of meson-side transverse-momentum and relativistic effects, and a dedicated numerical study based on a systematic comparison of different GTMD modeling strategies, are left for future investigation.

The remainder of this paper is organized as follows. Section~\ref{s:def} defines the relevant GTMDs and their relation to OAM. Section~\ref{s:results} outlines the theoretical framework for exclusive vector meson production and presents the derivation of azimuthal-angle-dependent observables. Finally, Sec.~\ref{s:conclusion} summarizes our findings and discusses the experimental implications.

%%%%%%%%%%%%%%%%%%%%%%%%%%%%%%%%%%%%%%%%%%%%%%%%%%%%% 
\section{Generalized TMDs of gluons} 
\label{s:def}
%%%%%%%%%%%%%%%%%%%%%%%%%%%%%%%%%%%%%%%%%%%%%%%%%%%%% 
The gluonic structure of the nucleon is parameterized by GTMDs. A systematic classification of gluon GTMDs for a spin-$\tfrac{1}{2}$ target was established in Ref.~\cite{Lorce:2013pza}. These distributions are defined through the off-forward gluon–gluon correlator,
%%%%%%%%
\begin{equation}
W_{\lambda,\lambda'}^{g\,[ij]}(P,\Delta,x, k_\perp)
=
\frac{1}{P^+}
\int \frac{dz^-\, d^2z_\perp}{(2\pi)^3}\,
e^{ik\cdot z}\,
\big\langle p',\lambda' \big|
F_a^{+i}\!\left(-\tfrac{z}{2}\right)
\,\mathcal W_{ab}\!\left(-\tfrac{z}{2},\tfrac{z}{2}\right)
F_b^{+j}\!\left(\tfrac{z}{2}\right)
\big| p,\lambda \big\rangle
\Big|_{z^+=0},
\label{e:gtmd_corr}
\end{equation}
%%%%%%%%
where the gluon fields are represented by the field-strength tensor $F_a^{\mu\nu}$ with color index $a$. Gauge invariance of this nonlocal operator is ensured by the Wilson line $\mathcal W_{ab}$, which connects the field operators along a path determined by the specific process (e.g., future-pointing for SIDIS, past-pointing for Drell-Yan). In the context of exclusive heavy meson production, the Wilson lines takes the future-pointing staple-like shape.

We restrict ourselves to the leading-twist sector, for which the indices
$i,j$ are transverse and the plus component refers to the light-cone direction.
Throughout, light-cone coordinates are defined as
$a^\pm=(a^0\pm a^3)/\sqrt{2}$ and $ a_\perp=(a^1,a^2)$.
The initial and final nucleon states are characterized by momenta
$p$ and $p'$ and helicities $\lambda$ and $\lambda'$, respectively.
It is convenient to introduce the average nucleon momentum $P = (p + p')/2$ and the momentum transfer $\Delta = p' - p$. The gluon's longitudinal momentum fraction and transverse momentum are denoted by $x = k^+/P^+$ and $k_\perp$, respectively. The skewness parameter $\xi = -\Delta^+/(2P^+)$ characterizes the longitudinal momentum transfer between the initial and final nucleon states.

At leading twist, the correlator admits a decomposition in terms of
16 independent gluon GTMDs~\cite{Lorce:2013pza}.
In the present work we focus on the projections
%%%%%%%%
\begin{eqnarray}
W_{\lambda,\lambda'}^g
=
\delta_\perp^{ij}\, W_{\lambda,\lambda'}^{g\,[ij]},
\qquad
\widetilde W_{\lambda,\lambda'}^g
=
-\, i\,\varepsilon_\perp^{ij}\, W_{\lambda,\lambda'}^{g\,[ij]},
\end{eqnarray}
%%%%%%%%
which correspond to unpolarized and helicity-dependent gluon configurations,
respectively. The tensors $\delta_\perp^{ij}=-g_\perp^{ij}$ and
$\varepsilon_\perp^{ij}=\varepsilon^{-+ij}$ are defined with
$\varepsilon^{0123}=1$.
The correlator $W_{\lambda,\lambda'}^g$ can be parameterized in close analogy to the unpolarized quark case~\cite{Bhattacharya:2018lgm},
%%%%%%%%
\begin{eqnarray}
W_{\lambda,\lambda'}^g
& = & \frac{1}{2M} \, \bar{u}(p',\lambda') \bigg[ 
F_{1,1}^g + \frac{i  \sigma^{i+}  k_\perp^i}{P^+} \, F_{1,2}^g + \frac{i \, \sigma^{i+} \Delta_\perp^i}{P^+} \, F_{1,3}^g 
+ \frac{i \sigma^{ij} k_{\perp}^i \Delta_{\perp}^j}{M^2} \, F_{1,4}^g  \bigg] u(p,\lambda)
\nonumber \\
& = & \frac{1}{M \sqrt{1 - \xi^2}} \bigg\{ 
\bigg[ M \delta_{\lambda,\lambda'} - \frac{1}{2} \Big( \lambda \Delta_\perp^1 + i \Delta_\perp^2 \Big) \delta_{\lambda,-\lambda'} \bigg] F_{1,1}^g
+ (1 - \xi^2) \Big( \lambda k_\perp^1 + i k_\perp^2 \Big) \delta_{\lambda,-\lambda'} \, F_{1,2}^g
\nonumber \\
& & \hspace{0.8cm} 
+ \; (1 - \xi^2) \Big( \lambda \Delta_\perp^1 + i \Delta_\perp^2 \Big) \delta_{\lambda,-\lambda'} \, F_{1,3}^g
+ \frac{i \varepsilon_\perp^{ij} k_{\perp}^i \Delta_{\perp}^j}{M^2} \bigg[ \lambda M \delta_{\lambda,\lambda'} - \frac{\xi}{2} \Big( \Delta_\perp^1 + i \lambda \Delta_\perp^2 \Big) \delta_{\lambda,-\lambda'} \bigg] F_{1,4}^g \bigg\} \,,
\label{e:GTMD_unpol}
\end{eqnarray}
%%%%%%%%
while $\widetilde W_{\lambda,\lambda'}^g$ admits a decomposition analogous to
that of longitudinally polarized quarks~\cite{Bhattacharya:2018lgm},
%%%%%%%%
\begin{eqnarray}
\widetilde{W}_{\lambda, \lambda'}^g & = & \frac{1}{2M} \, \bar{u}(p',\lambda') \bigg[ 
- \frac{i \varepsilon_\perp^{ij} k_{\perp}^i \Delta_{\perp}^j}{M^2} \, G_{1,1}^g
+ \frac{i  \sigma^{i+}  \gamma_5 k_\perp^i}{P^+} \, G_{1,2}^g + \frac{i  \sigma^{i+}  \gamma_5 \Delta_\perp^i}{P^+} \, G_{1,3}^g
+ i \sigma^{+-} \gamma_5 \, G_{1,4}^g  \bigg] u(p,\lambda)
\nonumber \\
& = & \frac{1}{M \sqrt{1 - \xi^2}} \bigg\{
- \frac{i \varepsilon_\perp^{ij} k_{\perp}^i \Delta_{\perp}^j}{M^2} \bigg[ M \delta_{\lambda,\lambda'} - \frac{1}{2} \Big( \lambda \Delta_\perp^1 + i \Delta_\perp^2 \Big) \delta_{\lambda,-\lambda'} \bigg] G_{1,1}^g
+ (1 - \xi^2) \Big( k_\perp^1 + i \lambda k_\perp^2 \Big) \delta_{\lambda,-\lambda'} \, G_{1,2}^g
\nonumber \\
& & \hspace{0.8cm} 
+ \; (1 - \xi^2) \Big( \Delta_\perp^1 + i \lambda \Delta_\perp^2 \Big) \delta_{\lambda,-\lambda'} \, G_{1,3}^g
+ \bigg[ \lambda M \delta_{\lambda,\lambda'} - \frac{\xi}{2} \Big( \Delta_\perp^1 + i \lambda \Delta_\perp^2 \Big) \delta_{\lambda,-\lambda'} \bigg] G_{1,4}^g \bigg\} \,.
\label{e:GTMD_hel}
\end{eqnarray}
%%%%%%%%
(An additional set of eight GTMDs is required to describe linearly polarized gluons~\cite{Lorce:2013pza}; we do not consider them here, as they give rise to angular structures different from those of interest in this work.)
The parametrizations above closely parallel those used for quark GTMDs,
and our notation follows the conventions introduced in
Ref.~\cite{Meissner:2009ww}.
The nucleon spinors $u(p,\lambda)$ and $u(p',\lambda')$ are taken to be light-cone helicity spinors~\cite{Soper:1972xc,Diehl:2003ny}, and the nucleon mass is denoted by $M$.
A generic gluon GTMD
$X^g=X^g(x,\xi, k_\perp^2, \Delta_\perp^2, {k}_\perp \cdot {\Delta}_\perp)$ combines the dependence
on intrinsic transverse momentum familiar from TMDs with the off-forward
kinematics characteristic of GPDs.
In general, these distributions are complex-valued functions
of their arguments~\cite{Meissner:2008ay,Meissner:2009ww}.
For simplicity, we do not display their scale dependence explicitly;
see Ref.~\cite{Echevarria:2016mrc} for a detailed discussion.  More studies related to GTMDs can be found in Refs.~\cite{Tan:2023vvi,Manley:2024pcl,Bertone:2025vgy,Yang:2025neu,Benic:2025xpg,Lorce:2025aqp,Tan:2025gho,Chakrabarti:2025qba,Yu:2024ovn,Hagiwara:2021xkf,Vary:2025yqo,Chakrabarti:2025qba,Lorce:2025aqp,Puhan:2025kz,Sharma:2024arf,Tan:2024dmz}.

Among the various GTMDs, $F_{1,4}^g$ and $G_{1,1}^g$ are unique as they represent genuine Wigner-type correlations that do not reduce to any conventional GPDs or TMDs. These functions occupy a distinct sector of nucleon structure, vanishing upon integration over either the partonic transverse momentum $k_\perp$ or the nucleon recoil $\Delta_\perp$. Despite their elusive nature, they play a central role in probing the proton’s orbital and spin-orbit structure. The function $F_{1,4}^g$ encodes the distortion of the unpolarized gluon distribution in a longitudinally polarized nucleon; importantly, its transverse-momentum integral directly yields the canonical gluon orbital angular momentum density $L^g(x)$~\cite{Lorce:2011kd, Hatta:2011ku}:
%%%%%%%%
\begin{equation}
x L^g (x, \xi) = - \int d^2 k_\perp \, \frac{{k}_\perp^2}{M^2} \, F_{1,4}^{g}(x, k_\perp, \xi, \Delta_\perp=0) \,.
\label{e:oam}
\end{equation}
%%%%%%%%
This relation enables a direct determination of the Jaffe-Manohar $L^g \equiv \int_0^1 dx\, L^g(x, \xi=0)$ from experimental data, provided that the $k_\perp$-weighted moment of $F_{1,4}^g$ can be measured.

The distribution $G_{1,1}^g$, by contrast, describes the distortion of the longitudinally polarized gluon distribution in an unpolarized nucleon, arising from the coupling between the gluon's helicity and its orbital motion. Its transverse-momentum moment is given by~\cite{Bhattacharya:2024sck,Bhattacharya:2024sno}
%%%%%%%%
\begin{equation}
x \, C^g(x, \xi) = \int d^2 k_\perp \, \frac{k_\perp^2}{M^2} \, G_{1,1}^g(x, k_\perp, \xi, \Delta_\perp=0) \, .
\label{e:so}
\end{equation}
%%%%%%%%
This relation enables a direct determination of the strength of the gluon spin--orbit correlation,
which we identify with $C^g(x,\xi=0)$.
A non-zero value for this indicates that the gluon helicity is correlated with its transverse motion, a quantum-mechanical effect with no classical analogue. Recent studies in the small-$x$ regime suggest that this correlation is maximal in the saturation limit, hinting at a universal feature of QCD matter at high energies~\cite{Bhattacharya:2024sno}.

Finally, we emphasize that in our exclusive heavy-meson production process, the observables we derive are directly proportional to the transverse moments of these GTMDs. Specifically, subleading twist-3 corrections to the scattering amplitude generate terms linear in $k_\perp$, which, upon integration over the phase space, select exactly the moments required to access $L^g$ and $C^g$ according to Eqs.~(\ref{e:oam}) and~(\ref{e:so}).

%%%%%%%%%%%%%%%%%%%%%%%%%%%%%%%%%%%%%%%%%%%%%%%%%%%%%   
\section{Theoretical framework and results} 
\label{s:results}
%%%%%%%%%%%%%%%%%%%%%%%%%%%%%%%%%%%%%%%%%%%%%%%%%%%%% 
%%%%%%%%%%%%%%%%%%%%%%%%
\subsection{Kinematics}
%%%%%%%%%%%%%%%%%%%%%%%%
We consider exclusive meson electroproduction in electron--proton scattering. In the one-photon exchange approximation, the relevant subprocess is
%%%%%%%%%%%%%%
\begin{equation}
\gamma^*(q) + p(p,\lambda) \;\rightarrow\; m(q') + p'(p',\lambda'), \quad m = V/AV
\end{equation}
%%%%%%%%%%%%%%
where $q$ is the virtual photon momentum, $p(p,\lambda)$ and $p'(p',\lambda')$ denote the initial and final nucleon states, respectively, and $q'$ is the momentum of the meson $m$, which can be either a vector ($V$) or an axial-vector ($AV$) meson.
The standard invariants are
$
Q^2=-q^2, \quad p^2=p'^2=M^2, \quad q'^2=m_m^2, \quad t=(p-p')^2, \quad W^2=(p+q)^2.
$
The skewness parameter $\xi$ is defined as
%%%%%%%%%%%%%%
\begin{equation}
\xi = \frac{Q^2+m_m^2}{2W^2-2M^2+Q^2-m_m^2+t} 
\approx \frac{Q^2+m_m^2}{2W^2+Q^2-m_m^2},
\end{equation}
%%%%%%%%%%%%%%
where the approximation neglects $|t|\ll Q^2$ and nucleon-mass corrections\footnote{Throughout this work, we neglect nucleon-mass effects of order $M^2/Q^2$.}. It is convenient to introduce
$
\xi = \frac{\tilde{x}_B}{2-\tilde{x}_B}, \ 
\tilde{x}_B = \frac{Q^2+m_m^2}{W^2+Q^2},
$
which illustrates the relation between the skewness parameter $\xi$ and the generalized Bjorken variable. 
In the light-meson limit, $\tilde{x}_B \to x_B$ (where $x_B$ is the usual Bjorken variable), recovering the familiar expression, 
whereas for heavy quarkonium production, the threshold for the hard subprocess is shifted by the meson mass $m_m$.

The momenta of the virtual photon and the produced meson are parametrized as
$
q = \left(-\frac{Q^2}{2q^-},\, q^-,\, {0}_\perp\right), \
q' = \left(\frac{m_m^2+{\Delta}_\perp^2}{2q^-},\, q^-,\, -{\Delta}_\perp\right),
$
We now turn to the polarization vectors of the photon and meson. For longitudinal polarizations, we choose
%%%%%%%%%%%%%%
\begin{align}
\epsilon^{\gamma^*,L} &= \frac{1}{Q}\left(\frac{Q^2}{2q^-}, q^-, {0}_\perp\right), \\
\epsilon^{m,L} &= \frac{1}{m_m}\left(\frac{-m_m^2+{\Delta}_\perp^2}{2q^-}, q^-, -{\Delta}_\perp\right),
\end{align}
%%%%%%%%%%%%%%
satisfying the transversality conditions $\epsilon^{\gamma^*,L}\cdot q = 0$ and $\epsilon^{m,L} \cdot q' = 0$, as well as the normalizations, $\epsilon^{\gamma^*,L} \cdot \epsilon^{\gamma^*,L} = +1$ and $\epsilon^{m,L} \cdot\epsilon^{m,L} = -1$.
For transverse polarizations we choose 
%%%%%%%%%%%%%%
\begin{align}
\epsilon^{\gamma^*,T} &= (0,0,{\epsilon}_{\perp}^{\gamma^*}), \\
\epsilon^{m,T} &= \left(-\frac{{\epsilon}_{\perp}^{m} \cdot {\Delta}_\perp}{q^-}, 0, {\epsilon}_{\perp}^{m}\right),
\label{e:ward}
\end{align}
%%%%%%%%%%%%%%
which satisfy the transversality conditions and standard normalizations, $\epsilon^{\gamma^*, T} \cdot \epsilon^{\gamma^*, T} = - ({\epsilon}_{\perp}^{\gamma^*})^2$ and
$\epsilon^{m,T} \cdot \epsilon^{m,T} = - ({\epsilon}_{\perp}^{m})^2$. Here, the label $T$ denotes transverse polarization, while the additional subscript $\perp$ is reserved to indicate the transverse components of the corresponding vectors. Although alternative parametrizations of $\epsilon^{m,T}$ with a nonzero minus component are possible, we adopt the above form for definiteness.

%%%%%%%%%%%%%%%%%%%%%%%%
\subsection{Scattering amplitude}
%%%%%%%%%%%%%%%%%%%%%%%%
%%%%%%%%%%%%%
\begin{figure}[t]
    \centering
    \includegraphics[width = 10cm]{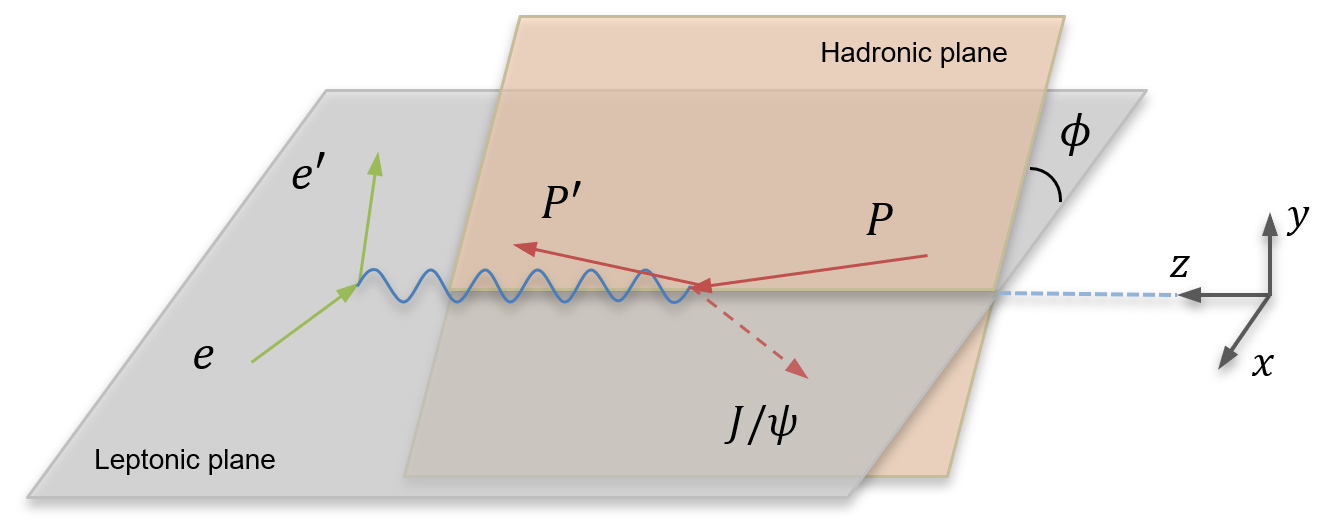}
    \caption{Schematic illustration of the exclusive heavy-meson production process. The figure is intended to provide an intuitive overview of the process. The $J/\psi$ has been labeled as a representative case of a vector meson.}
    \label{f1}
\end{figure}
%%%%%%%%%%%%%
Since we consider heavy-meson production, the heavy quark--antiquark pair is not
sourced from the nucleon wave function but is generated perturbatively. The hard
scattering therefore proceeds dominantly through gluon exchange at leading
order (LO).
At this order, the production amplitudes can be constructed straightforwardly from the six Feynman diagrams. In this section, we show Fig.~\ref{f1}, which provides a schematic illustration of the overall scattering process, while Fig.~\ref{f2} (see Appendix) collects the detailed Feynman diagrams entering the LO calculation. 
The nonperturbative nucleon matrix element is parameterized in terms of the $k_\perp$-moment of the gluon GTMD or a gluon GPD through the application of the collinear twist expansion, as discussed below. On the other hand, the hadronization of the heavy quark--antiquark pair into the final-state meson is described by the corresponding meson distribution amplitude, whose explicit vertex structure depends on whether the produced meson is a vector or an axial-vector state and is specified below.

The scattering amplitude $\mathcal{M}$ can be represented schematically in the following factorized form~\footnote{A fully transverse-momentum-dependent treatment of the meson side would more naturally be formulated in terms of a meson light-front wave function rather than a purely collinear distribution amplitude. In particular, for light vector meson production, phenomenological approaches often retain the quark transverse momentum in the meson wave function and in the hard subprocess, together with Sudakov suppression~\cite{Goloskokov:2005sd,Goloskokov:2006hr,Favart:2015umi}. In the present work, however, our primary aim is to isolate the nucleon-side kinematic twist-3 mechanism that gives access to the relevant gluon GTMD moments. For the heavy-quarkonium channel emphasized in the main text, we therefore restrict ourselves to the standard leading approximation consistent with collinear factorization combined with NRQCD~\cite{Collins:1997fj,Ivanov:2004vd,Chen:2019ysy,Flett:2021wfe}. A more complete treatment of meson-side transverse-momentum and relativistic corrections is left for future work~\cite{Lappi:2020ufv}.}:
%%%%%%%%%%%%%%
\begin{equation}
\mathcal{M} \propto \int dx \, \int dz \, \int d^2k_\perp \, H(x, \xi, z, k_\perp, \Delta_\perp) f^g(x, \xi, k_\perp, \Delta_\perp) \, \phi_m(z),
\label{e:hard1}
\end{equation}
%%%%%%%%%%%%%%
where $H$ denotes the perturbatively calculable hard-scattering kernel,
$f^g$ represents the nonperturbative proton matrix element, parameterized in
terms of either the $k_\perp$-moment of  GTMDs or GPDs, and $\phi_m$ denotes the leading meson-side nonperturbative input (the collinear distribution amplitude describing the longitudinal momentum-fraction distribution of the quark–antiquark pair inside the meson, with $z$ denoting the fraction carried by the quark) in the present approximation. We stress that Eq.~(\ref{e:hard1}) is intended here as a schematic representation adapted to our nucleon-side collinear twist expansion, rather than as the most general transverse-momentum-dependent factorization formula for the full process. (See, for instance, Refs.~\cite{Copeland:2023wbu,Copeland:2023qed,Copeland:2025vop} in the context of gluon TMD studies.)
To systematically separate contributions of different twist, the hard kernel
$H$ is expanded in powers of the transverse momenta ${k}_\perp$ and
${\Delta}_\perp$ relative to the hard scale of the process, which in our case is the photon virtuality $Q^2$. This procedure constitutes the
collinear twist expansion, originally developed in
Refs.~\cite{Ellis:1982cd,Ellis:1982wd}. Truncating the expansion at twist~3, one
finds 
%%%%%%%%%%%%%%
\begin{equation}
H(k_\perp, \Delta_\perp) \approx H(k_\perp = 0, \Delta_\perp = 0) + \frac{\partial H(k_\perp, \Delta_\perp=0)}{\partial k_\perp^\mu}\Big|_{k_\perp = 0} k_\perp^\mu + \frac{\partial H(k_\perp = 0, \Delta_\perp)}{\partial \Delta_\perp^\mu}\Big|_{\Delta_\perp = 0} \Delta_\perp^\mu  . 
\label{e:hard2}
\end{equation}
%%%%%%%%%%%%%%
In terms of twist counting, the first term corresponds to the leading-twist
(twist-2) contribution, while the second and third terms represent subleading
twist-3 corrections. Substituting Eq.~(\ref{e:hard2}) into
Eq.~(\ref{e:hard1}) and performing the ${k}_\perp$ integration, the term
linear in ${k}_\perp$ gives access to transverse-momentum--weighted moments
of GTMDs, whereas the term linear in ${\Delta}_\perp$ probes GPDs. It is the
former contribution that encodes sensitivity to partonic orbital angular
momentum and spin--orbit correlations, as explicitly demonstrated in
Eqs.~(\ref{e:oam})--(\ref{e:so}).

Returning to the effective vertex rules for vector and axial-vector mesons, we
summarize the corresponding replacements of the quark spinors. For a vector quarkonium state, the quark--antiquark spinor bilinear can be
replaced as~\cite{Ivanov:2004vd,Braaten:2002fi}
%%%%%%%%%%%%%%
$
v_i(q_2)\,\bar u_j(q_1)
\;\rightarrow\;
\frac{\delta_{ij}}{4N_c}
\left(\frac{\langle O_1\rangle_V}{m_q}\right)^{1/2}
\slashed{\epsilon}^{\,*}\,(\slashed{q}'+m_V),
$
where $q_1=q_2=q'/2$, with $q'$ denoting the four-momentum of the quarkonium
state, and $\epsilon^\mu$ the polarization vector of the vector meson. The
indices $i$ and $j$ label color in the fundamental representation, $N_c$ is the
number of colors, and the factor $\delta_{ij}$ projects the heavy quark--antiquark
pair onto a color-singlet state. The quantity $\langle O_1\rangle_V$ (and similarly
$\langle O_1\rangle_A$ for the axial-vector case discussed below) denotes the
leading nonrelativistic QCD (NRQCD) long-distance matrix element governing
quarkonium formation.
We also employ the nonrelativistic approximation
$m_M = 2m_q$. This choice is consistent with the standard heavy-quarkonium treatment in which the meson side is described at leading order in the velocity expansion by the appropriate NRQCD long-distance matrix element. In this sense, the meson-side structure retained in the present analysis corresponds to the leading heavy-quarkonium approximation, while more refined light-front wave functions and relativistic corrections lie beyond the scope of this work~\cite{Ivanov:2004vd,Chen:2019ysy,Flett:2021wfe}.
For an axial-vector meson, the corresponding replacement takes the form
$
v_i(q_2)\,\bar u_j(q_1)
\;\rightarrow\;
\frac{\delta_{ij}}{4N_c}
\left(\frac{\langle O_1\rangle_A}{m_q}\right)^{1/2}
\left(\slashed{q}_2-\frac{1}{2} m_{AV}\right)
\slashed{\epsilon}^{\,*}\gamma_5
\left(\slashed{q}_1+\frac{1}{2} m_{AV}\right),
$
where $q_1 = z\,q'$ and $q_2 = (1-z)\,q'$.
We find that the axial-vector contribution vanishes at the symmetric point
$z=1/2$, which motivates presenting the corresponding vertex rule for
$z\neq 1/2$. The resulting expressions for the amplitude at generic values of
$z$ are, however, rather lengthy and cumbersome. For this reason, we focus
primarily on the vector-meson case in the main text, while providing
representative results for axial-vector meson production at general $z$ in
Appendix~\ref{app}.

We now present the scattering amplitudes for specific polarization configurations
of the virtual photon and the vector meson. We adopt the notation
$\mathcal{M}^{\gamma^*, M}$ for amplitudes associated with GTMDs (GPDs) that conserve the proton helicity, and
$\widetilde{\mathcal{M}}^{\gamma^*, M}$ for amplitudes associated with GTMDs (GPDs) that do not conserve the proton helicity. The results are:
%%%%%%%%%%%%%%%%%%%%%%%
\begin{eqnarray}
 {\cal M}^{LL}\!\!&=& C  \delta_{\lambda, \lambda'}    \left \{ {\cal  H}_{eff}^{LL}\right \} , \\[0.1cm]
%%%%
{\cal M}^{TT}\!\!&=&C \delta_{\lambda, \lambda'}  
  \left ({\epsilon}^V_{\perp} \cdot {\epsilon}^{\gamma*}_\perp \right )\left \{ {\cal H}_{eff}^{TT} \right \} ,
   \\[0.1cm]
%%%%
{\cal M}^{TL}\!\!&=& C \delta_{\lambda,\lambda'}  \left \{ \left({\epsilon}^{\gamma*}_\perp  \cdot {\Delta}_\perp\right)   \big ({\cal H}^{TL}_{eff} + {\cal F}^{TL}_{1,1} \big )+\lambda \left( {\epsilon}^{\gamma*}_\perp  \times {\Delta}_\perp \right ) {\cal F}_{1,4}^{TL} +\left({\epsilon}^{\gamma*}_\perp  \cdot {\Delta}_\perp \right)   {\cal G} _{1,1}^{TL} - \lambda \left({\epsilon}_{\perp}^{\gamma*} \times {\Delta}_{\perp}\right) \big ( \mathcal{H'}^{TL}_{eff}+\mathcal{G}_{1,4}^{TL}  \right \} ,  \\[0.1cm]
%%%%
{\cal M}^{LT}\!\!&=&0 ,
\end{eqnarray}
%%%%%%%%%%%%%%%%%%%%%%%
and
%%%%%%%%%%%%%%%%%%%%%%%
\begin{eqnarray}
 \widetilde {\cal M}^{LL}\!\!&=& C  \delta_{\lambda,-\lambda'}  \left \{ 
 \left ({\Delta}_\perp  \times {S}_\perp \right )\widetilde{\cal E} ^{LL} \right \} ,  \\[0.1cm]
%%%%
\widetilde {\cal M}^{TT}\!\!&=& C  \delta_{\lambda,-\lambda'}  ({\epsilon}_{\perp}^V \cdot {\epsilon}^{\gamma*}_\perp)\left \{ 
 \left ({\Delta}_\perp  \times {S}_\perp\right) \widetilde{\cal E}^{TT}\right \} ,
   \\[0.1cm]
%%%%
\widetilde {\cal M}^{TL}\!\!&=& C  \delta_{\lambda,-\lambda'}  \left \{ 
 \left ( {\epsilon}^{\gamma*}_\perp  \times {S}_\perp \right ) \widetilde{\cal F} _{1,2}^{TL} - \left ( {\epsilon}^{\gamma*}_\perp  \times {S}_\perp \right)  \widetilde{\cal G} _{1,2}^{TL}\right \} ,
 \\[0.1cm]
 %%%%
\widetilde {\cal M}^{LT}\!\!&=& 0 ,
 \end{eqnarray}
%%%%%%%%%%%%%%%%%%%%%%%
where 
%%%%%%%%%%%%%%%%%%%%%%%
\begin{equation}
C
=
\frac{16 \, i\, \pi \, e\, e_q\, \alpha_s f_{V} \, m_V}{N_c \, (m^2_V + Q^2)}, \qquad
f_{V}
\equiv
\left( \frac{2\,\langle O_1 \rangle_{V}}{m_{V}} \right)^{1/2}.
\label{e:c_vector}
\end{equation}
%%%%%%%%%%%%%%%%%%%%%%%

In the above equations, we have defined the Compton form factors of the GTMDs (GPDs) as detailed below.  
As before, we adopt the convention that Compton form factors associated with helicity-conserving transitions carry no tilde, whereas those associated with helicity-nonconserving transitions are denoted with a tilde:
%%%%%%%%%%%%%%%%%%%%%%%
\begin{align}
\mathcal{H}^{LL}_{eff} & = - \int_{-1}^{1}  \dfrac{dx}{(x-\xi + i \varepsilon)(x+\xi - i \varepsilon)} \dfrac{Q \sqrt{1-\xi^2}}{m_V} H_{eff}^g,  \\[0.1cm]
%%%%%
\mathcal{\widetilde{E}}^{LL} & = - \int_{-1}^{1}  \dfrac{dx}{(x-\xi + i \varepsilon)(x+\xi - i \varepsilon)} \dfrac{Q}{m_V} \dfrac{1}{2M \sqrt{1-\xi^2}} E^g , \\[0.1cm]
%%%%%
\mathcal{H}^{TT}_{eff} & = \int_{-1}^{1}  \dfrac{dx}{(x-\xi + i \varepsilon)(x+\xi - i \varepsilon)} \sqrt{1-\xi^2} H_{eff}^g, \\[0.1cm]
%%%%%
\mathcal{\widetilde{E}}^{TT} & = \int_{-1}^{1}  \dfrac{dx}{(x-\xi + i \varepsilon)(x+\xi - i \varepsilon)} \dfrac{1}{2M \sqrt{1-\xi^2}} E^g,  
%\\[0.1cm]
\end{align}
%%%%%
\begin{align}
\mathcal{H}^{TL}_{eff} & = \int_{-1}^{1}  \dfrac{dx}{(x-\xi + i \varepsilon)^2(x+\xi - i \varepsilon)^2} \dfrac{\xi^2 \sqrt{1-\xi^2}}{m_V} H_{eff}^g, \\[0.1cm]
%%%%%
\mathcal{H'}^{TL}_{eff} & =  \int_{-1}^{1}  \dfrac{dx}{(x-\xi + i \varepsilon)^2(x+\xi - i \varepsilon)^2}  \dfrac{i x \xi\sqrt{1-\xi^2}}{m_V} H^{'g}_{eff} ,   \\[0.1cm]
%%%%%
\mathcal{F}^{TL}_{1,1} & = \int_{-1}^{1}  \dfrac{dx}{(x-\xi - i \varepsilon)^2(x+\xi + i \varepsilon)^2}  \dfrac{2x\xi}{m_V} \dfrac{1}{\sqrt{1-\xi^2}} \int d^2k_\perp \bigg ( \dfrac{k_\perp \cdot \Delta_\perp}{\Delta^2_\perp} \bigg ) F_{1,1}^g,
\\[0.1cm]
%%%%%
{\cal G}_{1,1}^{TL} &= -\frac{1}{\sqrt{1-\xi^2}} \int_{-1}^{1} dx \frac{\xi^2}{m_V (x+\xi- i \varepsilon)^2 (x-\xi+ i \varepsilon)^2} \int d^2 k_\perp \, 
\left( \frac{k_\perp^2}{M^2} \right) G_{1,1}^g,  
\\[0.1cm]
%%%%%
\widetilde{\cal F}_{1,2}^{TL} &=  M \sqrt{1-\xi^2} \int_{-1}^{1} dx \frac{x \xi}{m_V (x+\xi- i \varepsilon)^2 (x-\xi+ i \varepsilon)^2} \int d^2 k_\perp \, 
\left( \frac{k_\perp^2}{M^2} \right) F_{1,2}^g, \\[0.1cm]
%%%%%
\widetilde{\cal G}_{1,2}^{TL} &= -M \sqrt{1-\xi^2} \int_{-1}^{1} dx \frac{\xi^2}{m_V (x+\xi- i \varepsilon)^2 (x-\xi+ i \varepsilon)^2} \int d^2 k_\perp \, 
\left( \frac{k_\perp^2}{M^2} \right) G_{1,2}^g, 
\\[0.1cm]
%%%%%
{\cal F}_{1,4}^{TL} &=  \frac{1}{\sqrt{1-\xi^2}} \int_{-1}^{1} dx \frac{i x \xi}{m_V (x+\xi- i \varepsilon)^2 (x-\xi+ i \varepsilon)^2} \int d^2 k_\perp \, 
\left( \frac{k_\perp^2}{M^2} \right) F_{1,4}^g ,\\[0.1cm]
%%%%%
\mathcal{G}^{TL}_{1,4} & = \int \dfrac{dx}{(x-\xi+ i \varepsilon)^2(x+\xi - i \varepsilon)^2}  \dfrac{2i \xi^2}{m_V} \dfrac{1}{\sqrt{1-\xi^2}} \int d^2k_\perp \bigg ( \dfrac{k_\perp \cdot \Delta_\perp}{\Delta^2_\perp} \bigg ) G_{1,4}^g \, .
\end{align}
%%%%%%%%%%%%%%%%%
As is evident, the amplitudes can be expressed as linear combinations of $k_\perp$ moments of GTMDs and GPDs themselves. 
In particular, contributions involving GPDs arise when the hard kernel is of twist~2 or of twist~3 and linear in $\Delta_\perp$, whereas contributions involving $k_\perp$ moments of GTMDs are entirely of twist~3, as discussed around Eq.~(\ref{e:hard2}). Here we use the shorthand notation (see Appendix~\ref{appa} for the explicit GPD parametrization):
\begin{align}
    H_{eff}^g \equiv H^g - \dfrac{\xi^2}{1-\xi^2} E^g \, ,\qquad
    H^{'g}_{eff} \equiv \widetilde{H}^g - \dfrac{\xi^2}{1-\xi^2} \widetilde{E}^g \, .
\end{align}
%%%%%%%%%%%%%%%%%%%%%%%
It should be noted that, to ensure QED gauge invariance, one must parametrize the transverse polarization of the meson very carefully: we find that neglecting the $+$ or $-$ components (see Eq.~(\ref{e:ward})) leads to a violation of the QED Ward identity. 
This subtlety emphasizes the importance of correctly treating the meson polarization in such calculations.

%%%%%%%%%%%%%%%%%%%%%%%%
\subsection{Cross section}
%%%%%%%%%%%%%%%%%%%%%%%%
In this section, we derive the explicit dependence of the differential cross-section on the azimuthal angle $\phi$, defined as the angle between the lepton scattering plane and the hadron production plane. The differential cross-section for the process $e + p \to e' + p' + V$ can be expressed in terms of the contraction of the leptonic tensor $L_{\mu\nu}$ with the hadronic tensor $ \mathcal{M}^\mu \mathcal{M}^{*\nu}$:
%%%%%%%%%%%%%%%%%%%
\begin{equation}
d\sigma = \frac{\mathcal{C}}{2s} \, L_{\mu\nu} \, \mathcal{M}^\mu \mathcal{M}^{*\nu},
\end{equation}
%%%%%%%%%%%%%%%%%%%
where $\mathcal{C}$ includes the flux and phase-space factors.
The leptonic tensor is given by
%%%%%%%%
\begin{equation}
L_{\mu\nu}
=
2\left(l_\mu l'_\nu + l_\nu l'_\mu\right)
-
g_{\mu\nu}\,Q^2 ,
\label{e:leptonten}
\end{equation}
%%%%%%%%
which follows from a straightforward calculation and is valid for unpolarized leptons, to which we restrict our analysis.

For the explicit evaluation of the cross section, it is convenient to work in the hadron frame. In order to isolate the azimuthal (\(\phi\)) dependence, we expand the bilinear
\(\mathcal{M}^\mu \mathcal{M}^{*\nu}\) in a complete basis of independent tensors. To this end, we introduce three mutually orthogonal four-vectors defined in the hadron frame,
%%%%%%%%
\begin{equation}
T^\mu = (1,0,0,0), \qquad
X^\mu = (0,1,0,0), \qquad
Y^\mu = (0,0,1,0),
\label{e:cartesian}
\end{equation}
%%%%%%%%
where we follow the Cartesian index convention \((0,1,2,3)\), rather than light-cone coordinates, which simplifies the calculations presented below.
Current conservation, 
%%%%%%%% 
\begin{equation} q_\mu (\mathcal{M}^\mu\, \mathcal{M}^{*\nu}) = q_\nu (\mathcal{M}^\mu\, \mathcal{M}^{*\nu}) = 0 \, .
\end{equation} 
%%%%%%%% 
This relation, which is analogous to $q_\mu W^{\mu\nu} = q_\nu W^{\mu\nu}$ in semi-inclusive processes (where $W^{\mu\nu}$ is the hadronic tensor), implies that the bilinear
\(\mathcal{M}^\mu \mathcal{M}^{*\nu}\) can be decomposed into nine independent tensor structures. We choose the following basis:
%%%%%%%%
\begin{align}
\mathcal{V}_1^{\mu\nu} &= X^\mu X^\nu + Y^\mu Y^\nu, \\
\mathcal{V}_2^{\mu\nu} &= T^\mu T^\nu, \\
\mathcal{V}_3^{\mu\nu} &= T^\mu X^\nu + X^\mu T^\nu, \\
\mathcal{V}_4^{\mu\nu} &= X^\mu X^\nu - Y^\mu Y^\nu, \\
\mathcal{V}_5^{\mu\nu} &= i\left(T^\mu X^\nu - X^\mu T^\nu\right), \\
\mathcal{V}_6^{\mu\nu} &= i\left(X^\mu Y^\nu - Y^\mu X^\nu\right), \\
\mathcal{V}_7^{\mu\nu} &= i\left(T^\mu Y^\nu - Y^\mu T^\nu\right), \\
\mathcal{V}_8^{\mu\nu} &= T^\mu Y^\nu + Y^\mu T^\nu, \\
\mathcal{V}_9^{\mu\nu} &= X^\mu Y^\nu + Y^\mu X^\nu .
\label{e:main1}
\end{align}
%%%%%%%%
The spirit of this decomposition follows Ref.~\cite{Koike:2003zc}, where a systematic classification was introduced for a semi-inclusive process. In the present work, we adopt the same approach and apply it to our exclusive process. The linear independence of the tensors \(\{\mathcal{V}_i^{\mu\nu}\}\) was explicitly verified by demonstrating that a vanishing linear combination, $\sum_{i=1}^{9} c_i\,\mathcal{V}_i^{\mu\nu} = 0$,
implies that all coefficients must vanish, $c_i = 0$ for all $i$.
Since we restrict ourselves to unpolarized electrons, the leptonic tensor contains only a symmetric part, as shown in Eq.~(\ref{e:leptonten}). Consequently, only the symmetric components of the bilinear $\mathcal{M}^\mu \mathcal{M}^{*\nu}$ can contribute to the cross section. This immediately restricts the relevant tensor structures to
$\mathcal{V}_1^{\mu \nu}$, $\mathcal{V}_2^{\mu \nu}$, $\mathcal{V}_3^{\mu \nu}$, $\mathcal{V}_4^{\mu \nu}$, $\mathcal{V}_8^{\mu \nu}$, and $\mathcal{V}_9^{\mu \nu}$,
while the remaining antisymmetric tensors do not contribute.
The expansion coefficients of $\mathcal{M}^\mu \mathcal{M}^{*\nu}$ in the basis $\{\mathcal{V}_i^{\mu\nu}\}$ are then obtained by introducing the corresponding inverse tensors $\widetilde{\mathcal{V}}_i^{\mu\nu}$ $(i=1,\ldots,9)$, leading to
%%%%%%%%%%%%
\begin{align}
{\cal M}^\mu {\cal M}^{*\nu}
=
\sum_{k=1}^{9}
{\cal V}_k^{\mu\nu}
\left[
\widetilde{{\cal V}}_k^{\rho\sigma} \, {\cal M}_\rho {\cal M}^{*}_\sigma
\right].
\label{e:expansion_coeff}
\end{align}
%%%%%%%%%%%%
The explicit forms of the inverse tensors are given by
%%%%%%%%%%%%
\begin{align}
\widetilde{{\cal V}}_1^{\mu\nu}
&=
\frac{1}{2}
\left(
X^\mu X^\nu + Y^\mu Y^\nu
\right),
\\
\widetilde{{\cal V}}_2^{\mu\nu}
&=
T^\mu T^\nu,
\\
\widetilde{{\cal V}}_3^{\mu\nu}
&=
-\frac{1}{2}
\left(
T^\mu X^\nu + X^\mu T^\nu
\right),
\\
\widetilde{{\cal V}}_4^{\mu\nu}
&=
\frac{1}{2}
\left(
X^\mu X^\nu - Y^\mu Y^\nu
\right),
\\
\widetilde{{\cal V}}_5^{\mu\nu}
&=
\frac{i}{2}
\left(
T^\mu X^\nu - X^\mu T^\nu
\right),
\\
\widetilde{{\cal V}}_6^{\mu\nu}
&=
-\frac{i}{2}
\left(
X^\mu Y^\nu - Y^\mu X^\nu
\right),
\\
\widetilde{{\cal V}}_7^{\mu\nu}
&=
\frac{i}{2}
\left(
T^\mu Y^\nu - Y^\mu T^\nu
\right),
\\
\widetilde{{\cal V}}_8^{\mu\nu}
&=
-\frac{1}{2}
\left(
T^\mu Y^\nu + Y^\mu T^\nu
\right),
\\
\widetilde{{\cal V}}_9^{\mu\nu}
&=
\frac{1}{2}
\left(
X^\mu Y^\nu + Y^\mu X^\nu
\right).
\end{align}
%%%%%%%%%%%%
Since only symmetric tensor structures contribute to the cross section for unpolarized electrons, the same restriction applies to the inverse basis. Consequently, the nonvanishing inverse tensors associated with symmetric structures are
$\widetilde{\mathcal{V}}_1^{\mu\nu}$,
$\widetilde{\mathcal{V}}_2^{\mu\nu}$,
$\widetilde{\mathcal{V}}_3^{\mu\nu}$,
$\widetilde{\mathcal{V}}_4^{\mu\nu}$,
$\widetilde{\mathcal{V}}_8^{\mu\nu}$,
and
$\widetilde{\mathcal{V}}_9^{\mu\nu}$.

We now have all the ingredients necessary to decompose the cross section into the following two components:
%%%%%%%%
\begin{equation}
d\sigma = \frac{\mathcal{C}}{2s} \, \sum_{k=1}^{9}\left(L_{\mu\nu} \, 
{\cal V}_k^{\mu\nu}\right)
\left(
\widetilde{{\cal V}}_k^{\rho\sigma} \, {\cal M}_\rho {\cal M}^{*}_\sigma
\right).
\label{e:cross_master}
\end{equation}
%%%%%%%%
Here, we first explain the essence of this equation and outline our calculation strategy, before presenting the explicit results in the following paragraphs.
In Eq.~(\ref{e:cross_master}), the first term captures the azimuthal-angle dependence of the cross section, while the second term corresponds to different tensor structures contracted with the scattering amplitudes. To relate the latter to the amplitudes for differently polarized virtual photons, we express the basis vectors in $\mathcal{V}_i^{\mu\nu}$ in light-cone coordinates using virtual photon polarization vectors~\footnote{For notational simplicity, we drop the $\gamma^*$ index from the polarization tensors from this point onward.}:
%%%%%%%%%%%%%%%%%%%%%
\begin{equation}
\epsilon_0^\mu = \frac{1}{\sqrt{2}}(1,1,0,0), \quad
\epsilon_+^\mu = \frac{1}{\sqrt{2}}(0,0,-1,i), \quad
\epsilon_-^\mu = \frac{1}{\sqrt{2}}(0,0,1,i).
\end{equation}
%%%%%%%%%%%%%%%%%%%%%
Here, the index ``0'' denotes the longitudinal polarization of the virtual photon, while ``+'' and ``-'' label the transverse polarization states with positive- and negative-helicity, respectively.
Using these polarization vectors as building blocks, we redefine the three mutually orthogonal four-vectors introduced in Eq.~(\ref{e:cartesian}):
%%%%%%%%%%%%%%%%%%%%%
\begin{equation}
T^\mu = \epsilon_0^\mu, \quad
X^\mu = -\frac{1}{\sqrt{2}}\left(\epsilon_+^\mu - \epsilon_-^\mu\right), \quad
Y^\mu = -\frac{i}{\sqrt{2}}\left(\epsilon_+^\mu + \epsilon_-^\mu\right).
\label{e:lc}
\end{equation}
%%%%%%%%%%%%%%%%%%%%%
This allows us to express the content of the second set of parentheses in Eq.~(\ref{e:cross_master}) in terms of the $\gamma^* p$ amplitude. For completeness, we have explicitly verified that the resulting tensor basis in Eq.~(\ref{e:lc}) remains linearly independent when expressed in terms of these photon polarization vectors. The same checks were performed for the inverse tensors, yielding similar results.

Let us now explicitly present the results of Eq.~(\ref{e:cross_master}) following the strategies outlined above.  
To this end, we consider a reference frame in which the virtual-photon four-momentum, written in the light-cone basis, is parametrized as
%%%%%%%%%%%%%%%%%%%%%
\begin{equation}
q^\mu = \left(-\frac{Q}{\sqrt{2}}, \frac{Q}{\sqrt{2}}, {0}_\perp \right).
\end{equation}
%%%%%%%%%%%%%%%%%%%%%
In this frame, the incoming lepton momentum can be written as
%%%%%%%%%%%%%%%%%%%%%
\begin{equation}
l^\mu = \left(
\frac{Q(1-y)}{\sqrt{2}\,y}, \frac{Q}{\sqrt{2}\,y}, 
\frac{Q\sqrt{1-y}}{y} \cos \phi_{l_\perp}, \frac{Q\sqrt{1-y}}{y} \sin \phi_{l_\perp} \right) \, ,
\end{equation}
%%%%%%%%%%%%%%%%%%%%%
where $y \equiv \frac{p \cdot q}{p \cdot l}$ is the usual lepton energy fraction and $\phi_{l_\perp}$ is the azimuthal angle of the lepton transverse momentum in the lepton plane.
The outgoing lepton momentum follows from momentum conservation.  
Upon contracting the resulting leptonic tensor with the tensor structures of the basis constructed from the photon polarization vectors, as described above, we obtain the following distinct angular modulations:
%%%%%%%%%%%%%%%%%%%%%
\begin{equation}
\begin{aligned}
L_{\mu\nu}\,\mathcal{V}_3^{\mu\nu} &\propto \cos\phi, &
L_{\mu\nu}\,\mathcal{V}_4^{\mu\nu} &\propto \cos 2\phi, \\
L_{\mu\nu}\,\mathcal{V}_8^{\mu\nu} &\propto \sin\phi, &
L_{\mu\nu}\,\mathcal{V}_9^{\mu\nu} &\propto \sin 2\phi.
\end{aligned}
\end{equation}
%%%%%%%%%%%%%%%%%%%%%
Here, $\phi$ may equivalently be identified as the angle between the lepton scattering plane and the hadron production plane provided we choose $\Delta_\perp = (|\vec{\Delta}_\perp|,\,0)$, which we adopt to keep the expressions simple (and for this reason we henceforth drop the subscript $l_\perp$ from the angle). These angular modulations are well known in the literature~\cite{Koike:2003zc,Goloskokov:2009ia,Goloskokov:2007nt} and originate from specific interference patterns: the $\cos\phi$ and $\sin\phi$ terms arise from the interference between twist-2 and twist-3 amplitudes, while the $\cos 2\phi$ and $\sin 2\phi$ terms arise from the interference between twist-3 amplitudes.

Next, we work, without loss of generality, with an initial proton of positive helicity and proceed to calculate the second term in parentheses of Eq.~(\ref{e:cross_master}).  
Below, the photon-helicity amplitudes will be denoted by ${\cal M}_{Vp',\gamma^* p}$.  
It then follows straightforwardly that the differential cross section for exclusive vector–meson electroproduction can be written as
%%%%%%%%%
\begin{align}
\dfrac{d \sigma}{dy dQ^2 dt d \phi} & =\dfrac{\alpha_{\rm em}}{2 \pi^2} \dfrac{\left(1-y+\dfrac{y^2}{2}\right)}{yQ^2} \bigg [ \frac{d\sigma_T}{dt} + \epsilon \frac{d\sigma_L}{dt}  +\sqrt{2\epsilon(1+\epsilon)} \cos \phi \frac{d\sigma_{LT}}{dt} + \, \epsilon \cos 2\phi \, \frac{d\sigma_{TT}}{dt}\nonumber \\ 
& + \sqrt{2\epsilon(1+\epsilon)} \sin \phi \frac{d\sigma_{\rm sin \phi}}{dt} +\, \epsilon \sin 2\phi \, \frac{d\sigma_{\rm sin 2\phi}}{dt} \bigg ]
\label{eq:ep_cross_section}
\end{align}
%%%%%%%%%
where the individual structure functions are expressed as
%%%%%%%%%
\begin{align}
\frac{d\sigma_T}{dt}
&=
\dfrac{x^2_B}{32 \pi Q^4}   
\left(
|{\cal M}_{Vp',++}|^2
+
|{\cal M}_{Vp',-+}|^2
\right),
\\[0.15cm]
\frac{d\sigma_L}{dt}
&=
\dfrac{x^2_B}{16 \pi Q^4}   
|{\cal M}_{Vp',0+}|^2,
\\[0.15cm]
\frac{d\sigma_{LT}}{dt}
&=
\dfrac{\sqrt{2} x^2_B}{32\pi Q^4}
\operatorname{Re}
\!\left[
{\cal M}_{Vp',0+}
\left(
{\cal M}_{Vp',-+}^*
-
{\cal M}_{Vp',++}^*
\right)
\right],
\\[0.15cm]
\frac{d\sigma_{TT}}{dt}
&=
-\dfrac{x^2_B}{16 \pi Q^4}  
\operatorname{Re}
\!\left[
{\cal M}_{Vp',++}\,{\cal M}_{Vp',-+}^*
\right],
\\[0.15cm]
\frac{d\sigma_{\rm sin \phi}}{dt}
&=
\dfrac{\sqrt{2}x^2_B}{32\pi Q^4}
\operatorname{Im}
\!\left[
{\cal M}_{Vp',0+}^*
\left(
{\cal M}_{Vp',-+}
+
{\cal M}_{Vp',++}
\right)
\right],
\\[0.15cm]
\frac{d\sigma_{\rm sin 2\phi}}{dt}
&=
\dfrac{x^2_B}{16 \pi Q^4}  
\operatorname{Im}
\!\left[
{\cal M}_{Vp',++}\,{\cal M}_{Vp',-+}^*
\right].
%\label{e:main1}
\end{align}
%%%%%%%%%
Here we have expressed our results in terms of the virtual-photon flux parameter $\epsilon$, defined as, $
\epsilon
=
\frac{1 - y}{1 - y + \frac{1}{2} y^2}$. 
The first four terms in Eq.~\eqref{eq:ep_cross_section} represent the standard parity-even structure functions, which have been extensively studied in the context of leading-twist GPD phenomenology~\cite{Diehl:2003ny, Goloskokov:2008ib}. Specifically, $d\sigma_T$ and $d\sigma_L$ denote the cross sections for transversely and longitudinally polarized virtual photons, respectively. The modulations $(\cos\phi, \sin\phi)$ and $(\cos 2\phi, \sin 2\phi)$ arise from the interference between longitudinal and transverse photon helicity amplitudes, and between transverse photon helicity amplitudes of opposite helicity, respectively. The $\cos$-type terms correspond to the real part of the interference, while the $\sin$-type terms correspond to the imaginary part.
Our results for these terms are in full agreement with the existing literature; see also Appendix~\ref{appa}, where we explicitly take the GPD limit of our results and recover the known results.

We now present explicit results for the structure functions for each meson polarization.
In the case of a vector meson with positive circular polarization ($V=+$), the results read:
%%%%%%%%%%%
\begin{align}
\frac{d\sigma_T^+}{dt}
&= \frac{x_B^2}{32 \pi Q^4} \, C^2
\left\{
\delta_{\lambda,\lambda'} \left|{\cal H}_{eff}^{TT}\right|^2
+ \delta_{\lambda,-\lambda'} \, \left|\vec{\Delta}_\perp\right|^2
 \left|{\cal \widetilde{E}}^{TT}\right|^2
\right\} ,
\end{align}
%%%%%%%%%%%
and vanishing in all other cases. Similarly, in the case of a vector meson with negative circular polarization ($V=-$), the results read:
%%%%%%%%%%%
\begin{align}
\frac{d\sigma_T^-}{dt}
&= \frac{x_B^2}{32 \pi Q^4} \, C^2
\left\{
\delta_{\lambda,\lambda'} \left|{\cal H}_{eff}^{TT}\right|^2
+ \delta_{\lambda,-\lambda'} \, \left|\vec{\Delta}_\perp\right|^2
 \left|{\cal \widetilde{E}}^{TT}\right|^2
\right\} \, ,
\end{align}
%%%%%%%%%%%
and vanishing in all other cases. Both of the above results arise from the interference between twist-2 contributions. Thus, we find that for circularly polarized vector mesons, the cross section is sensitive only to GPDs.

Finally, for longitudinally polarized vector meson ($V=0$), the results read:
%%%%%%%%%%%
\begin{align}
\frac{d\sigma_T^0}{dt}
&= \frac{x_B^2}{32 \pi Q^4} \, C^2
\Bigg\{
\delta_{\lambda,\lambda'}
\Big[
|\vec{\Delta}_\perp|^2
\left| {\cal H}_{eff}^{TL} + {\cal F}_{1,1}^{TL} + {\cal G}_{1,1}^{TL}\right|^2
+ |\vec{\Delta}_\perp|^2
\left|{\cal F}_{1,4}^{TL} - {\cal H'}_{eff}^{TL} - {\cal G}_{1,4}^{TL}\right|^2
\Big] \nonumber \\
& \hspace{4cm} + 2 \, \delta_{\lambda,-\lambda'}
\left|\widetilde{\cal F}_{1,2}^{TL} - \widetilde{\cal G}_{1,2}^{TL}\right|^2
\Bigg\},
\\[0.6em]
%%%%%%%%%%%
\frac{d\sigma_L^0}{dt}
&= \frac{x_B^2}{16 \pi Q^4} \, C^2
\Bigg\{
\delta_{\lambda,\lambda'}
\left|{\cal H}_{eff}^{LL}\right|^2
+ \delta_{\lambda,-\lambda'}
|\vec{\Delta}_\perp|^2
\left|\widetilde{\cal E}^{LL}\right|^2
\Bigg\},
\\[0.6em]
%%%%%%%%%%%
\frac{d\sigma_{LT}^0}{dt}
&= \frac{\sqrt{2}\, x_B^2}{32 \pi Q^4} \, C^2
\Bigg\{
\delta_{\lambda,\lambda'} \,
\sqrt{2} |\vec{\Delta}_\perp|\,
\operatorname{Re}\!\Big[
{\cal H}_{eff}^{LL}
\left({\cal H}_{eff}^{TL*} + {\cal F}_{1,1}^{TL*}
+ {\cal G}_{1,1}^{TL*}\right)
\Big]
\nonumber \\
&  \hspace{4cm}+ \delta_{\lambda,-\lambda'} \,
\sqrt{2} |\vec{\Delta}_\perp|\,
\operatorname{Re}\!\Big[
\mathcal{\widetilde{E}}^{LL}
\left(\widetilde{\cal F}_{1,2}^{TL*}
- \widetilde{\cal G}_{1,2}^{TL*}\right)
\Big]
\Bigg\},
%\\[0.6em]
\end{align}
%%%%%%%%%%%
\begin{align}
\frac{d\sigma_{TT}^0}{dt}
&= -\frac{x_B^2}{16 \pi Q^4} \, C^2
\Bigg\{
\delta_{\lambda,\lambda'}
\Big[
-\dfrac{1}{2} |\vec{\Delta}_\perp|^2
\left|{\cal H}_{eff}^{TL}+{\cal F}_{1,1}^{TL}
+ {\cal G}_{1,1}^{TL}\right|^2
+ \dfrac{1}{2} |\vec{\Delta}_\perp|^2
\left|{\cal F}_{1,4}^{TL} - {\cal H'}_{eff}^{TL}
- {\cal G}_{1,4}^{TL}\right|^2
\Big]
\Bigg\},
\\[0.6em]
%%%%%%%%%%%
\frac{d\sigma_{\rm sin\phi}^0}{dt}
&= \frac{\sqrt{2}\, x_B^2}{32 \pi Q^4} \, C^2
\Bigg\{
- \delta_{\lambda,\lambda'} \, \lambda
|\vec{\Delta}_\perp|\,
\operatorname{Re}\!\Big[
\sqrt{2}\, {\cal H}_{eff}^{LL*}
\left({\cal F}_{1,4}^{TL} - {\cal H'}_{eff}^{TL}
- {\cal G}_{1,4}^{TL}\right)
\Big]
\nonumber \\
& \hspace{4cm}- \lambda \, \delta_{\lambda,-\lambda'}
|\vec{\Delta}_\perp|\,
\operatorname{Im}\!\Big[ \sqrt{2} \,
\widetilde{\mathcal{E}}^{LL*}
\left(\widetilde{\cal F}_{1,2}^{TL}
- \widetilde{\cal G}_{1,2}^{TL}\right)
\Big]
\Bigg\},
\\[0.6em]
%%%%%%%%%%%
\frac{d\sigma_{\rm sin2\phi}^0}{dt}
&= \frac{x_B^2}{16 \pi Q^4} \, C^2
\Bigg\{
- \delta_{\lambda, \lambda'}\lambda |\vec{\Delta}_\perp|^2\,
\operatorname{Re}\!\Big[
\left( {\cal H}_{eff}^{TL}+ {\cal F}_{1,1}^{TL}
+ {\cal G}_{1,1}^{TL}\right)
\left({\cal F}_{1,4}^{TL*} - {\cal H'}_{eff}^{TL*}
- {\cal G}_{1,4}^{TL*}\right)
\Big]
\Bigg\}.
\end{align}
%%%%%%%%%%%
Here, $d\sigma^0_L/dt$ arises from twist-2 × twist-2 contributions, while the $d\sigma^0_{LT}/dt$ and $d\sigma^0_{\rm sin\phi}/dt$ terms result from the interference between twist-2 and twist-3 contributions. All remaining terms involve purely twist-3 × twist-3 contributions.  
From these expressions, we conclude that the polarization-independent $\cos 2\phi$ and polarization-dependent $\sin 2\phi$ terms provide exceptionally clean signatures of the gluon GTMDs $F_{1,4}$ and $G_{1,1}$. While other observables, such as $d\sigma_T/dt$ and $d\sigma_{\rm sin\phi}/dt$, are also sensitive to these GTMDs, they receive contributions from multiple distributions and are thus less selective. 
Furthermore, the $\cos 2\phi$ modulation stands out as a particularly striking signal: it can be measured with an unpolarized proton, and unlike in processes such as diffractive dijet production, it carries no information about linear gluon polarization in this exclusive color-singlet channel. Consequently, this signal provides a clean and well-isolated laboratory signal for probing the canonical gluon orbital angular momentum and gluon spin–orbit correlations. Note that for the sine-type terms, our results show that these contributions are always accompanied by a helicity factor $\lambda$, which is required to ensure parity conservation in the strong interaction.

The above results present the cross sections and their polarization-dependent contributions for the vector meson.  In principle, the total cross section summed over all vector meson polarizations can be obtained by adding the contributions from each polarization state. For a generic cross section labeled by $X \in \{T, L, LT, TT, \sin\phi, \sin 2\phi\}$, one may write:
%%%%%%%%%%%%%%
\begin{align}
    \frac{d\sigma_X}{dt} 
    = \frac{d\sigma_X^+}{dt} + \frac{d\sigma_X^-}{dt} + \frac{d\sigma_X^0}{dt}.
\end{align}
%%%%%%%%%%%%%%
This provides a straightforward way to combine the contributions from different meson polarizations, which may be useful in experimental analyses where one sums over the polarization states.

For the reasons outlined above, the experimental isolation of these twist-3 observables at the EIC appears promising. Although twist-3 effects are formally suppressed by factors of $\mathcal{O}(\langle k_\perp \rangle/Q)$ or $\mathcal{O}(\langle \Delta_\perp \rangle/Q)$ relative to the leading-twist cross section, they offer a significantly higher-rate channel compared to other proposed processes, such as diffractive dijet production or double Drell–Yan. We expect that the high luminosity of the EIC, together with the large acceptance of next-generation detectors like ePIC and the use of forward Roman Pots for recoil proton tagging, will enable the precise multi-dimensional binning necessary to extract these observables. We conclude by noting that a meaningful numerical analysis of the proposed asymmetries requires nonperturbative inputs for gluon GTMDs, which are not yet uniquely constrained by data. At present, one mainly has exploratory modeling strategies, including double-distribution-inspired constructions and related Radon-transform-based representations, as well as light-front spectator-model approaches with explicit gluonic degrees of freedom~\cite{Radyushkin:1998es,Musatov:1999xp,Anikin:2019xgh,Chakrabarti:2023dxa,Tan:2023djt,Chakrabarti:2024kjw,Chakrabarti:2025qba,Tan:2025xxx}. Existing studies also indicate that the resulting phenomenology can be quite sensitive to the modeling assumptions~\cite{Bhattacharya:2024qol,Boer:2023txn}.  We therefore defer a dedicated numerical study to future work, where different modeling strategies can be systematically compared and theoretical uncertainties assessed in a controlled manner.

%%%%%%%%%%%%%%%%%%%%%%%%%%%%%%%%%%%%%%%%%%%%%%%%%%%%%   
\section{Conclusions} 
\label{s:conclusion}
%%%%%%%%%%%%%%%%%%%%%%%%%%%%%%%%%%%%%%%%%%%%%%%%%%%%%
In this work, we have presented a systematic derivation of the azimuthal-angle--dependent cross section for exclusive (axial-)vector meson electroproduction within the framework of collinear twist-3 factorization. By analyzing the interference between different virtual-photon polarization states, we identified distinctive angular modulations in the cross section that provide direct sensitivity to specific gluon GTMDs. 
For clarity, the main body of the paper focuses on vector meson production. Owing to the increased complexity of the axial-vector case, the corresponding results are presented in a dedicated appendix.

For vector mesons, we demonstrated that the polarization-independent $\cos 2\phi$ and polarization-dependent $\sin 2\phi$ modulations, which arise from twist-3--twist-3 interference contributions, are sensitive to specific $k_\perp$-moments of the gluon GTMDs $F_{1,4}^g$ and $G_{1,1}^g$.
These distributions encode fundamental information on canonical gluon orbital angular momentum and partonic spin--orbit correlations. This sector of nucleon structure has no counterpart in the conventional generalized parton distribution or transverse-momentum--dependent frameworks, rendering the observables proposed here uniquely valuable probes of gluonic dynamics.

A notable advantage of the exclusive (axial-)vector meson channel is its comparatively high production rate relative to other proposed probes of gluon GTMDs, such as diffractive dijet production or exclusive double Drell--Yan processes. Although the observables identified here are of twist-3 origin, they remain experimentally promising due to the high luminosity and large acceptance anticipated at future facilities. In particular, the EIC, combined with forward proton tagging via Roman Pots and next-generation detectors such as ePIC, will enable the multi-dimensional binning required to extract these azimuthal asymmetries with precision.

Looking ahead, several important extensions of the present work are warranted. These include the incorporation of genuine twist-3 effects arising from tri-gluon correlation functions, as well as a more complete treatment of meson-side transverse-momentum and relativistic effects, which will be necessary for a fully comprehensive and quantitatively precise description of the process. In addition, we plan to carry out detailed numerical calculations and phenomenological studies tailored to the kinematics of both the EIC and the Electron--Ion Collider in China (EicC). A comprehensive phenomenological analysis, however, requires a systematic comparison of different GTMD modeling approaches together with a controlled assessment of the associated theoretical uncertainties, since gluon GTMDs currently remain largely unconstrained and are described primarily through exploratory modeling strategies~\cite{Radyushkin:1998es,Musatov:1999xp,Anikin:2019xgh,Chakrabarti:2023dxa,Tan:2023djt,Chakrabarti:2024kjw,Tan:2025xxx,Bhattacharya:2024qol,Boer:2023txn}. We therefore leave such a dedicated numerical study to future work.
While $J/\psi$ production provides a particularly clean and compelling signature at the EIC, its yield at the lower center-of-mass energies of the EicC is expected to be more limited. Nevertheless, the formalism developed here can be straightforwardly extended to light vector meson production, such as $\rho^0$, where a sufficiently large photon virtuality $Q^2$ can provide the hard scale required for the applicability of perturbative QCD. Such studies, together with systematic model comparisons and possible refinements on the meson/heavy-quarkonium side, will further clarify the experimental feasibility of mapping gluonic orbital angular momentum and spin--orbit correlations through exclusive processes.

%%%%%%%%%%%%%%%%%%%%
\begin{acknowledgments} 
D.~D. acknowledges support from a Summer 2025 predoctoral fellowship awarded by the Department of Physics at the University of Connecticut. This work has been supported by  the National Natural Science Foundation of China under Grant  No.12175118, No. 12505102~(D. ~Z.)
and No. 12321005~(J.~Z.) ,  the Shandong
Province Natural Science Foundation No. ZR2024QA104~(D. ~Z.).
\end{acknowledgments}
%%%%%%%%%%%%%%%%%%%%

%\clearpage
%%%%%%%%%%%%%%%%%%%%%%%%%%%%%%%%%%%%%%%%%%%%%%%%%%%%%   
\appendix
%%%%%%%%%%%%%%%%%%%%%%%%%%%%%%%%%%%%%%%%%%%%%%%%%%%%% 
\section{Amplitude for axial-vector meson production up to twist-3 accuracy}
\label{app}
%%%%%%%%%%%%%%%%%%%%%%%%%%%%%%%%%%%%%%%%%%%%%%%%%%%%% 
%%%%%%%%%%%%%
\begin{figure}[t]
    \centering
    \includegraphics[scale=0.48]{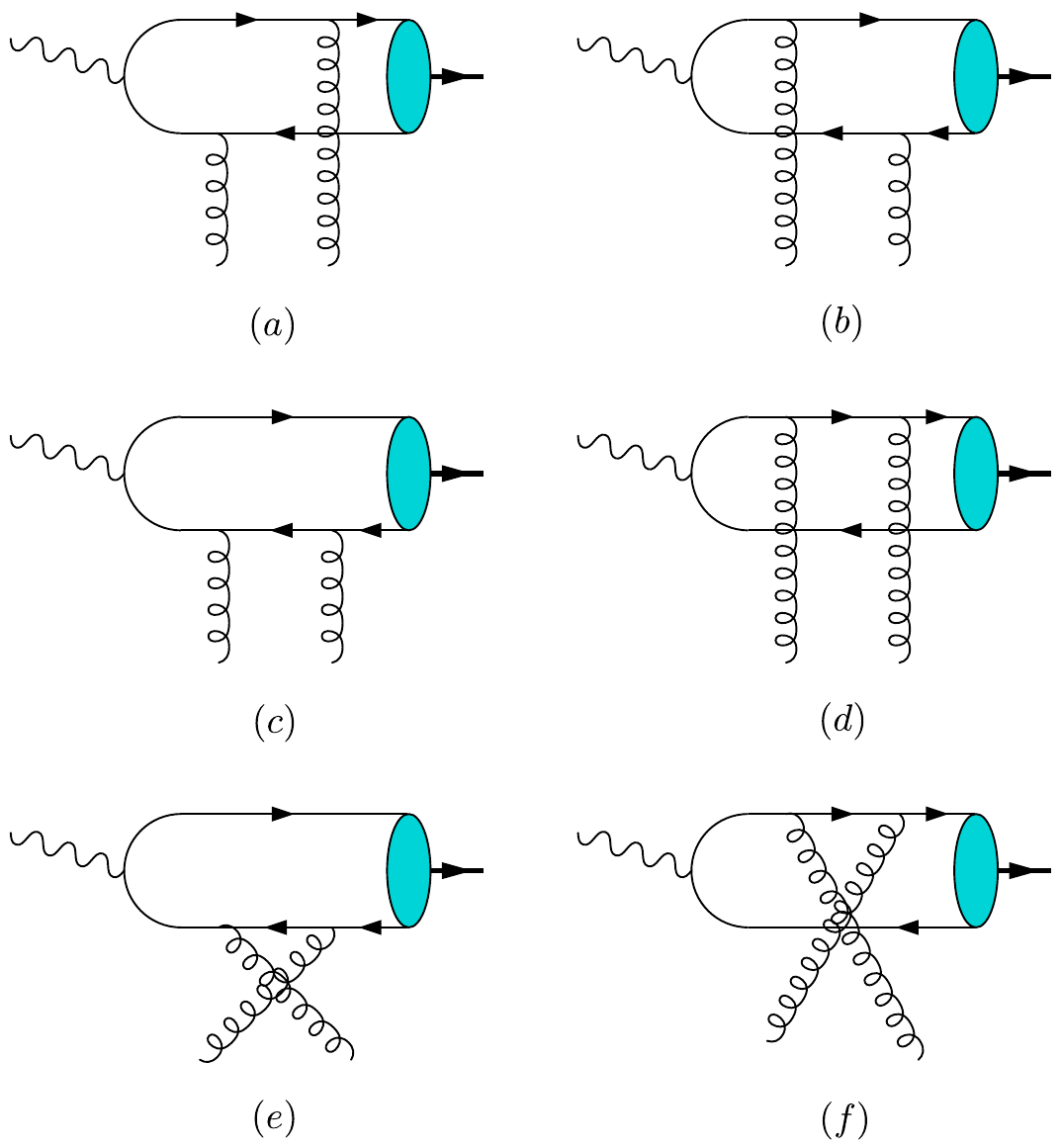}
    \caption{The six leading-order subprocesses contributing to the amplitude.} 
    \label{f2}
\end{figure}
%%%%%%%%%%%%%
In the following, we present the scattering amplitudes for axial-vector meson
production, evaluated diagram by diagram (see Fig.~\ref{f2}), including both
twist-2 and twist-3 contributions. The resulting expressions are, in general,
rather involved. However, both the twist-2 and twist-3 contributions vanish at
the symmetric point $z = \tfrac{1}{2}$. Consequently, obtaining a nonvanishing
result requires working away from this point, i.e., for $z \neq \tfrac{1}{2}$.

The amplitudes admit a common schematic structure. For the $F$-type GTMDs, involving $W^{g}$,  we write
%%%%%%%%%%%%
\begin{align}
{\cal M}
=
C_{AV} \int_{-1}^{1} dx \,
\frac{1}{(x-\xi + i \varepsilon)(x+\xi - i \varepsilon)}
\int_{0}^{1} dz \,
\int d^2 k_\perp \,
{\cal H}(x,\xi,z,k_\perp,\Delta_\perp)\,
W^{g}(x,\xi,k_\perp,\Delta_\perp)\,
\phi_{AV}(z),
\end{align}
%%%%%%%%%%%%
where ${\cal H}$ denotes the hard-scattering kernel. For the $G$-type GTMDs,
involving $\widetilde{W}^{g}$, the corresponding amplitudes are obtained by
placing tildes on the hard kernel, ${\cal H} \to \widetilde{\cal H}$.
The overall normalization factor $C_{AV}$ and the axial-vector meson decay constant $f_{AV}$
are given by
%%%%%%%%%%%%
\begin{align}
C_{AV}
=
\frac{i\, \pi \, e\, e_q\, \alpha_s f_{AV}}{2 N_c}, \qquad
f_{AV}
\equiv
\left( \frac{2\,\langle O_1 \rangle_{AV}}{m_{AV}} \right)^{1/2}.
\end{align}
%%%%%%%%%%%%

We now briefly describe how the results are organized below. For each diagram
and for each of the $F$- and $G$-type GTMDs, we explicitly quote the corresponding
hard-scattering kernels ${\cal H}$ or $\widetilde{\cal H}$. These kernels are classified according to the polarization configurations of the
virtual photon and the meson, as well as the twist order, which we denote
schematically by ${\cal H}^{\gamma^*, M}_{twist}$ or
${\cal \widetilde{H}}^{\gamma^*, M}_{twist}$.
The superscripts specify the photon and meson polarizations, while the subscripts
indicate the twist order, with the twist-2 ($t2$)  and twist-3 contributions arising from
$k_\perp$ ($t3,k_\perp$) and $\Delta_\perp$ ($t3,\Delta_\perp$) expansions presented separately.

%%%%%%%%%%%%%%%%%%%%%%%%%%%%%%%%%%%%%%%%%%%%%%%%%%%%%  
\subsection*{Diagram \texorpdfstring{$\boldsymbol{(a)}$}{(a)}}
%%%%%%%%%%%%%%%%%%%%%%%%%%%%%%%%%%%%%%%%%%%%%%%%%%%%% 
For diagram~($a$), we find that, for the $F$-type GTMDs, the case of a longitudinally
polarized virtual photon and a longitudinally polarized meson yields a vanishing
contribution at both twist-2 and twist-3.
%%%%%%%%%%%%
\begin{align}
{\cal H}^{LL}_{t2}=0, \quad {\cal H}^{LL}_{t3, k_\perp} =0, \quad {\cal H}^{LL}_{t3, \Delta_\perp} =0.
\end{align}
%%%%%%%%%%%%
We next present the remaining (non-zero) results:
%%%%%%%%%%%%
\begin{align}
{\cal H}^{TL}_{t3, k_\perp} & = -\frac{8 i m_{AV} \left(m^2_{AV}-Q^2\right) \xi^2 (1 -2  z)^2 \epsilon^{ij}_\perp (\epsilon^{\gamma* i}_\perp k_\perp^j)}{\left(m^2_{AV} \left(2 x (z-1)+\xi  \left(4 z^2-6 z+1\right)\right)+2 Q^2 (z-1) (\xi +x)\right) \left(m^2_{AV} \left(2 x z+\xi 
   \left(4 z^2-2 z-1\right)\right)+2 Q^2 z (x-\xi )\right)}\\
%%%
{\cal H}^{TL}_{t3, \Delta_\perp}   & = \frac{4 i m_{AV} \left(m^2_{AV}-Q^2\right) \xi ^2 (2 z-1)^3 \epsilon^{ij}_\perp (\epsilon^{\gamma* i}_\perp \Delta_\perp^j) }{\left(m^2_{AV} \left(2 x (z-1)+\xi  \left(4 z^2-6 z+1\right)\right)+2 Q^2 (z-1) (\xi +x)\right) \left(m^2_{AV} \left(2 x z+\xi 
   \left(4 z^2-2 z-1\right)\right)+2 Q^2 z (x-\xi )\right)}
\end{align}
%%%%%%%%%%%%
\begin{align}
{\cal H}^{LT}_{t3, \Delta_\perp}  &=  \frac{16 i m^2_{AV} \xi ^2 Q (2 z-1) \epsilon^{ij}_\perp (\epsilon^{AV i}_\perp \Delta_\perp^j) }{\left(m^2_{AV} \left(2 x
   (z-1)+\xi  \left(4 z^2-6 z+1\right)\right)+2 Q^2 (z-1) (\xi +x)\right) \left(m^2_{AV} \left(2 x z+\xi  \left(4 z^2-2 z-1\right)\right)+2
   Q^2 z (x-\xi )\right)}
\end{align}
%%%%%%%%%%%%
We find that, in the case of $F$-type GTMDs, the transversely polarized photon–meson channel gives no contribution at both twist-2 and twist-3:
%%%%%%%%%%%%
\begin{align}
{\cal H}^{TT}_{t2}=0, \quad {\cal H}^{TT}_{t3, k_\perp} =0, \quad {\cal H}^{TT}_{t3, \Delta_\perp} =0.
\end{align}
%%%%%%%%%%%%

We now turn to the results for the G-type GTMDs. The results that are non-zero are:
%%%%%%%%%%%%
\begin{align}
&{\cal \widetilde{H}}^{LL}_{t2} \nonumber \\
& = -\frac{2 m_{AV} (2 z-1) \left(m^2_{AV}+Q^2\right) \left(m^2_{AV} \left(x^2 (2 z-1)+2 \xi  x \left(4 z^2-4 z-1\right)+2 \xi ^2
   \left(4 z^3-6 z^2+1\right)\right)+Q^2 (2 z-1) \left(x^2-\xi ^2\right)\right)}{Q \left(m^2_{AV} \left(\xi -2 x z-4 \xi  z^2+2 \xi 
   z\right)+2 Q^2 z (\xi -x)\right) \left(m^2_{AV} \left(2 x (z-1)+\xi  \left(4 z^2-6 z+1\right)\right)+2 Q^2 (z-1) (\xi +x)\right)} \\
%%%
&{\cal \widetilde{H}}^{TL}_{t3,k_\perp} = -\frac{8 m_{AV} \xi  (2 z-1)  \left(m^2_{AV} \left(x (2 z-1)+2 \xi 
   \left(4 z^2-4 z-1\right)\right)+Q^2 x (2 z-1)\right) \delta^{ij}_\perp (\epsilon^{\gamma* i}_\perp k^j_\perp )}{\left(m^2_{AV} \left(2 x (z-1)+\xi  \left(4 z^2-6 z+1\right)\right)+2 Q^2 (z-1) (\xi
   +x)\right) \left(m^2_{AV} \left(2 x z+\xi  \left(4 z^2-2 z-1\right)\right)+2 Q^2 z (x-\xi )\right)} \\
%%%
&{\cal \widetilde{H}}^{TL}_{t3,\Delta_\perp} = \frac{4 m_{AV} \xi  (1-2 z)^2  \left(m^2_{AV} \left(x (2 z-1)+2 \xi 
   \left(4 z^2-4 z-1\right)\right)+Q^2 x (2 z-1)\right) \delta^{ij}_\perp (\epsilon^{\gamma* i}_\perp \Delta^j_\perp)}{\left(m^2_{AV} \left(2 x (z-1)+\xi  \left(4 z^2-6 z+1\right)\right)+2 Q^2 (z-1) (\xi
   +x)\right) \left(m^2_{AV} \left(2 x z+\xi  \left(4 z^2-2 z-1\right)\right)+2 Q^2 z (x-\xi )\right)} \\
%%%
&{\cal \widetilde{H}}^{LT}_{t3,\Delta_\perp} =\frac{4 m^2_{AV} \xi  (2 z-1) \left(m^2_{AV} (2 x+\xi  (2 z-1))+2
   Q^2 \left(x+\xi  \left(-4 z^3+6 z^2-1\right)\right)\right) \delta^{ij}_\perp (\epsilon^{AV i}_\perp \Delta^j_\perp)}{Q \left(m^2_{AV} \left(\xi -2 x z-4 \xi  z^2+2 \xi  z\right)+2 Q^2 z (\xi
   -x)\right) \left(m^2_{AV} \left(2 x (z-1)+\xi  \left(4 z^2-6 z+1\right)\right)+2 Q^2 (z-1) (\xi +x)\right)}
\end{align}
%%%%%%%%%%%%
We find that, in the case of $G$-type GTMDs, the transversely polarized photon–meson channel gives no contribution at both twist-2 and twist-3:
%%%%%%%%%%%%
\begin{align}
{\cal \widetilde{H}}^{TT}_{t2}=0, \quad {\cal \widetilde{H}}^{TT}_{t3, k_\perp} =0, \quad {\cal \widetilde{H}}^{TT}_{t3, \Delta_\perp}=0 .
\end{align}
%%%%%%%%%%%%

%%%%%%%%%%%%%%%%%%%%%%%%%%%%%%%%%%%%%%%%%%%%%%%%%%%%%  
\subsection*{Diagram \texorpdfstring{$\boldsymbol{(b)}$}{(b)}}
%%%%%%%%%%%%%%%%%%%%%%%%%%%%%%%%%%%%%%%%%%%%%%%%%%%%% 
For diagram~($b$), we find that, for the $F$-type GTMDs, the case of a longitudinally
polarized virtual photon and a longitudinally polarized meson yields a vanishing
contribution at both twist-2 and twist-3.
%%%%%%%%%%%%
\begin{align}
{\cal H}^{LL}_{t2}=0, \quad {\cal H}^{LL}_{t3, k_\perp} =0, \quad {\cal H}^{LL}_{t3, \Delta_\perp} =0.
\end{align}
%%%%%%%%%%%%
We next present the remaining (non-zero) results:
%%%%%%%%%%%%
\begin{align}
{\cal H}^{TL}_{t3, k_\perp} & = \frac{8 i m_{AV} \left(m^2_{AV}-Q^2\right) \xi^2(1 -2  z)^2 \epsilon^{ij}_\perp (\epsilon^{\gamma* i}_\perp k_\perp^j) }{\left(m^2_{AV} \left(\xi +2 x z-4 \xi  z^2+2 \xi  z\right)+2 Q^2 z (\xi +x)\right) \left(m^2_{AV} \left(2 x (z-1)+\xi  \left(-4
   z^2+6 z-1\right)\right)+2 Q^2 (z-1) (x-\xi )\right)} \\
%%%
{\cal H}^{TL}_{t3, \Delta_\perp}   & = \frac{4 i m_{AV} \xi ^2 (2 z-1)^3  (m_{AV}-Q) (m_{AV}+Q) \epsilon^{ij}_\perp (\epsilon^{\gamma* i}_\perp \Delta_\perp^j) }{\left(m^2_{AV} \left(\xi +2 x z-4 \xi  z^2+2 \xi  z\right)+2 Q^2 z (\xi +x)\right) \left(m^2_{AV} \left(2 x (z-1)+\xi  \left(-4
   z^2+6 z-1\right)\right)+2 Q^2 (z-1) (x-\xi )\right)} \\
%%%
{\cal H}^{LT}_{t3, \Delta_\perp}  &= \frac{16 i m^2_{AV} \xi ^2 Q (2 z-1) \epsilon^{ij}_\perp (\epsilon^{AV i}_\perp \Delta_\perp^j) }{\left(m^2_{AV} \left(\xi +2
   x z-4 \xi  z^2+2 \xi  z\right)+2 Q^2 z (\xi +x)\right) \left(m^2_{AV} \left(2 x (z-1)+\xi  \left(-4 z^2+6 z-1\right)\right)+2 Q^2 (z-1)
   (x-\xi )\right)}
\end{align}
%%%%%%%%%%%%
We find that, in the case of $F$-type GTMDs, the transversely polarized photon–meson channel gives no contribution at both twist-2 and twist-3:
%%%%%%%%%%%%
\begin{align}
{\cal H}^{TT}_{t2}=0, \quad {\cal H}^{TT}_{t3, k_\perp} =0, \quad {\cal H}^{TT}_{t3, \Delta_\perp} =0.
\end{align}
%%%%%%%%%%%%

We now turn to the results for the G-type GTMDs. The results that are non-zero are:
%%%%%%%%%%%%
\begin{align}
& {\cal \widetilde{H}}^{LL}_{t2} \nonumber \\
& = \frac{2 m_{AV} (2 z-1)  \left(m^2_{AV}+Q^2\right) \left(m^2_{AV} \left(x^2 (2 z-1)+2 \xi  x \left(-4 z^2+4 z+1\right)+2 \xi ^2
   \left(4 z^3-6 z^2+1\right)\right)+Q^2 (2 z-1) \left(x^2-\xi ^2\right)\right)}{Q \left(m^2_{AV} \left(\xi +2 x z-4 \xi  z^2+2 \xi 
   z\right)+2 Q^2 z (\xi +x)\right) \left(m^2_{AV} \left(\xi -2 x z+2 x+4 \xi  z^2-6 \xi  z\right)-2 Q^2 (z-1) (x-\xi )\right)} \\
%%%
& {\cal \widetilde{H}}^{TL}_{t3,k_\perp} = -\frac{8 m_{AV} \xi  (2 z-1)   \left(m^2_{AV} \left(-2 x z+x+2 \xi 
   \left(4 z^2-4 z-1\right)\right)+Q^2 x (1-2 z)\right) \delta^{ij}_\perp (\epsilon^{\gamma* i}_\perp k^j_\perp )}{\left(m^2_{AV} \left(\xi +2 x z-4 \xi  z^2+2 \xi  z\right)+2 Q^2 z (\xi +x)\right)
   \left(m^2_{AV} \left(2 x (z-1)+\xi  \left(-4 z^2+6 z-1\right)\right)+2 Q^2 (z-1) (x-\xi )\right)} \\
%%%
& {\cal \widetilde{H}}^{TL}_{t3,\Delta_\perp} =- \frac{4 m_{AV} \xi  (1-2 z)^2 \left(m^2_{AV} \left(-2 x z+x+2 \xi 
   \left(4 z^2-4 z-1\right)\right)+Q^2 x (1-2 z)\right) \delta^{ij}_\perp (\epsilon^{\gamma* i}_\perp \Delta^j_\perp )}{\left(m^2_{AV} \left(\xi +2 x z-4 \xi  z^2+2 \xi  z\right)+2 Q^2 z (\xi +x)\right)
   \left(m^2_{AV} \left(2 x (z-1)+\xi  \left(-4 z^2+6 z-1\right)\right)+2 Q^2 (z-1) (x-\xi )\right)} \\
%%%
& {\cal \widetilde{H}}^{LT}_{t3,\Delta_\perp} = -\frac{4 m^2_{AV} \xi  (2 z-1) \left(-\left(m^2_{AV} (\xi +2 x-2 \xi 
   z)\right)-2 Q^2 \left(\xi +x+4 \xi  z^3-6 \xi  z^2\right)\right) \delta^{ij}_\perp (\epsilon^{AV i}_\perp \Delta^j_\perp)}{Q \left(m^2_{AV} \left(\xi +2 x z-4 \xi  z^2+2 \xi  z\right)+2 Q^2 z (\xi
   +x)\right) \left(m^2_{AV} \left(\xi -2 x z+2 x+4 \xi  z^2-6 \xi  z\right)-2 Q^2 (z-1) (x-\xi )\right)}
\end{align}
%%%%%%%%%%%%
We find that, in the case of $G$-type GTMDs, the transversely polarized photon–meson channel gives no contribution at both twist-2 and twist-3:
%%%%%%%%%%%%
\begin{align}
{\cal \widetilde{H}}^{TT}_{t2}=0, \quad {\cal \widetilde{H}}^{TT}_{t3, k_\perp} =0, \quad {\cal \widetilde{H}}^{TT}_{t3, \Delta_\perp}=0 .
\end{align}
%%%%%%%%%%%%

%%%%%%%%%%%%%%%%%%%%%%%%%%%%%%%%%%%%%%%%%%%%%%%%%%%%%  
\subsection*{Diagram \texorpdfstring{$\boldsymbol{(c)}$}{(c)}}
%%%%%%%%%%%%%%%%%%%%%%%%%%%%%%%%%%%%%%%%%%%%%%%%%%%%% 
For diagram~($c$), we find that, for the $F$-type GTMDs, the case of a longitudinally
polarized virtual photon and a longitudinally polarized meson yields a vanishing
contribution at both twist-2 and twist-3.
%%%%%%%%%%%%
\begin{align}
{\cal H}^{LL}_{t2}=0, \quad {\cal H}^{LL}_{t3, k_\perp} =0, \quad {\cal H}^{LL}_{t3, \Delta_\perp} =0.
\end{align}
%%%%%%%%%%%%
We next present the remaining (non-zero) results:
%%%%%%%%%%%%
\begin{align}
& {\cal H}^{TL}_{t3, \Delta_\perp} = -\frac{4 i m_{AV} (2 z-1) \left(m^2_{AV} \left(-2 x z+x+\xi  \left(8
   z^3-8 z^2-4 z+1\right)\right)-Q^2 (2 z-1) (\xi +x-2 \xi  z)\right) \epsilon^{ij}_\perp (\epsilon^{\gamma* i}_\perp \Delta_\perp^j) }{\left(m^2_{AV} \left(4 z^2-4 z-1\right)+4 Q^2 (z-1)\right)
   \left(m^2_{AV} \left(\xi -2 x z+2 x+4 \xi  z^2-6 \xi  z\right)-2 Q^2 (z-1) (x-\xi )\right)} \\
 %%%
& {\cal H}^{LT}_{t3, \Delta_\perp} \nonumber \\
& = -\frac{4 i m^2_{AV} (2 z-1) \left(Q^2 \left(x \left(-2
   z^2+z+1\right)-\xi  \left(4 z^3-8 z^2+z-1\right)\right)-m^2_{AV} \left(2 z^2-z-1\right) (\xi +x-2 \xi  z)\right) \epsilon^{ij}_\perp (\epsilon^{AV i}_\perp \Delta_\perp^j) }{Q \left(m^2_{AV}
   \left(4 z^2-4 z-1\right)+4 Q^2 (z-1)\right) \left(m^2_{AV} \left(2 x (z-1)+\xi  \left(-4 z^2+6 z-1\right)\right)+2 Q^2 (z-1) (x-\xi
   )\right)} 
   \end{align}
   \begin{align}
&{\cal H}^{TT}_{t2} = -\frac{4 i m^2_{AV} z (2 z - 3) (2 z - 1)\left(m^2_{AV}+Q^2\right) (\xi +x) \epsilon^{ij}_\perp (\epsilon^{AV i}_\perp \epsilon^{\gamma* j}_\perp) }{\left(m^2_{AV} \left(4 z^2-4 z-1\right)+4 Q^2 (z-1)\right) \left(m^2_{AV} \left(\xi -2
   x z+2 x+4 \xi  z^2-6 \xi  z\right)-2 Q^2 (z-1) (x-\xi )\right)}
\end{align}
%%%%%%%%%%%%

We now turn to the results for the G-type GTMDs. The results that are non-zero are:
%%%%%%%%%%%%
\begin{align}
& {\cal \widetilde{H}}^{LL}_{t2} = \frac{2 m_{AV} (2 z-1)  \left(m^2_{AV}+Q^2\right) \left(-\left(m^2_{AV} \left(2 x+\xi  \left(-8 z^3+12 z^2-2
   z+1\right)\right)\right)-2 Q^2 \left(2 z^2-3 z+1\right) (x-\xi )\right)}{Q \left(m^2_{AV} \left(4 z^2-4 z-1\right)+4 Q^2 (z-1)\right)
   \left(m^2_{AV} \left(2 x (z-1)+\xi  \left(-4 z^2+6 z-1\right)\right)+2 Q^2 (z-1) (x-\xi )\right)} \\
%%%
& {\cal \widetilde{H}}^{TL}_{t3,\Delta_\perp} =- \frac{4 m_{AV} (2 z-1)   \left(-\left(m^2_{AV} \left(x (1-2 z)^2+\xi 
   \left(-8 z^3+12 z^2-2 z+1\right)\right)\right)-Q^2 \left(\xi +x (1-2 z)^2-2 \xi  z\right)\right) \delta^{ij}_\perp (\epsilon^{\gamma* i}_\perp \Delta^j_\perp )}{\left(m^2_{AV} \left(4 z^2-4 z-1\right)+4
   Q^2 (z-1)\right) \left(m^2_{AV} \left(\xi -2 x z+2 x+4 \xi  z^2-6 \xi  z\right)-2 Q^2 (z-1) (x-\xi )\right)} \\
%%%
& {\cal \widetilde{H}}^{LT}_{t3,\Delta_\perp} \nonumber \\
&= -\frac{4 m^2_{AV} (2 z-1) \left(m^2_{AV} \left(x \left(2
   z^2-z-1\right)+\xi  z \left(4 z^2-8 z+3\right)\right)+Q^2 \left(x \left(2 z^2-z-1\right)+\xi  \left(4 z^3-11 z+3\right)\right)\right) \delta^{ij}_\perp (\epsilon^{AV i}_\perp \Delta^j_\perp)}{Q
   \left(m^2_{AV} \left(4 z^2-4 z-1\right)+4 Q^2 (z-1)\right) \left(m^2_{AV} \left(\xi -2 x z+2 x+4 \xi  z^2-6 \xi  z\right)-2 Q^2 (z-1)
   (x-\xi )\right)} 
   \end{align}
%%%%%%%%%%%%%%%%%%%%%%
\begin{align}
& {\cal \widetilde{H}}^{TT}_{t2} = -\frac{4 m^2_{AV} z \left(4 z^2-8 z+3\right)  \left(m^2_{AV} +Q^2\right) (\xi +x) \delta^{ij}_\perp (\epsilon^{AV i}_\perp \epsilon^{\gamma* j}_\perp)}{\left(m^2_{AV}  \left(4 z^2-4 z-1\right)+4 Q^2 (z-1)\right) \left(m^2_{AV} \left(\xi -2
   x z+2 x+4 \xi  z^2-6 \xi  z\right)-2 Q^2 (z-1) (x-\xi )\right)}
\end{align}
%%%%%%%%%%%%

%%%%%%%%%%%%%%%%%%%%%%%%%%%%%%%%%%%%%%%%%%%%%%%%%%%%%  
\subsection*{Diagram \texorpdfstring{$\boldsymbol{(d)}$}{(d)}}
%%%%%%%%%%%%%%%%%%%%%%%%%%%%%%%%%%%%%%%%%%%%%%%%%%%%% 
For diagram~($d$), we find that, for the $F$-type GTMDs, the case of a longitudinally
polarized virtual photon and a longitudinally polarized meson yields a vanishing
contribution at both twist-2 and twist-3.
%%%%%%%%%%%%
\begin{align}
{\cal H}^{LL}_{t2}=0, \quad {\cal H}^{LL}_{t3, k_\perp} =0, \quad {\cal H}^{LL}_{t3, \Delta_\perp} =0.
\end{align}
%%%%%%%%%%%%
We next present the remaining results:
%%%%%%%%%%%%
\begin{align}
{\cal H}^{TL}_{t3, \Delta_\perp} & = \frac{4 i m_{AV} (2 z-1) \left(m^2_{AV} \left(-2 x z+x+\xi  \left(8
   z^3-16 z^2+4 z+3\right)\right)-Q^2 (2 z-1) (x+\xi  (2 z-1))\right) \epsilon^{ij}_\perp (\epsilon^{\gamma* i}_\perp \Delta_\perp^j) }{\left(m^2_{AV} \left(4 z^2-4 z-1\right)-4 Q^2 z\right)
   \left(m^2_{AV} \left(2 x z+\xi  \left(4 z^2-2 z-1\right)\right)+2 Q^2 z (x-\xi )\right)} \\
%%%
{\cal H}^{LT}_{t3, \Delta_\perp} & = \frac{4 i m^2_{AV} (2 z-1) \left(m^2_{AV} z (2 z-3) (x+\xi  (2
   z-1))+Q^2 \left(x z (2 z-3)+\xi  \left(-4 z^3+4 z^2+3 z-4\right)\right)\right) \epsilon^{ij}_\perp (\epsilon^{AV i}_\perp \Delta_\perp^j) }{Q \left(m^2_{AV} \left(-4 z^2+4 z+1\right)+4 Q^2 z\right)
   \left(m^2_{AV} \left(2 x z+\xi  \left(4 z^2-2 z-1\right)\right)+2 Q^2 z (x-\xi )\right)} \\
%%%
{\cal H}^{TT}_{t2} & = -\frac{4 i m^2_{AV} (2 z - 1) (2 z^2 - z - 1) \left(m^2_{AV}+Q^2\right) (\xi +x) \epsilon^{ij}_\perp (\epsilon^{AV i}_\perp \epsilon^{\gamma* j}_\perp) }{\left(m^2_{AV} \left(4 z^2-4 z-1\right)-4 Q^2 z\right) \left(m^2_{AV} \left(2 x z+\xi 
   \left(4 z^2-2 z-1\right)\right)+2 Q^2 z (x-\xi )\right)}   
\end{align}
%%%%%%%%%%%%

We now turn to the results for the G-type GTMDs. The results that are non-zero are:
%%%%%%%%%%%%
\begin{align}
& {\cal \widetilde{H}}^{LL}_{t2} = \frac{2 m_{AV} (2 z-1) \left(m^2_{AV}+Q^2\right) \left(m^2_{AV} \left(2 x+\xi  \left(8 z^3-12 z^2+2 z+3\right)\right)+2 Q^2 z (2
   z-1) (x-\xi )\right)}{Q \left(m^2_{AV} \left(-4 z^2+4 z+1\right)+4 Q^2 z\right) \left(m^2_{AV} \left(\xi -2 x z-4 \xi  z^2+2 \xi 
   z\right)+2 Q^2 z (\xi -x)\right)} \\
%%%
& {\cal \widetilde{H}}^{TL}_{t3,\Delta_\perp} = \frac{4 m_{AV} (2 z-1)  \left(m^2_{AV} \left(x (1-2 z)^2+\xi  \left(8
   z^3-12 z^2+2 z+3\right)\right)+Q^2 (2 z-1) (\xi +x (2 z-1))\right) \delta^{ij}_\perp (\epsilon^{\gamma* i}_\perp \Delta^j_\perp )}{\left(m^2_{AV} \left(4 z^2-4 z-1\right)-4 Q^2 z\right)
   \left(m^2_{AV} \left(2 x z+\xi  \left(4 z^2-2 z-1\right)\right)+2 Q^2 z (x-\xi )\right)} \\
%%%
& {\cal \widetilde{H}}^{LT}_{t3,\Delta_\perp} \nonumber \\
&=- \frac{4 m^2_{AV} (2 z-1) \left(m^2_{AV} \left(x (3-2 z) z+\xi 
   \left(4 z^3-4 z^2-z+1\right)\right)+Q^2 \left(x (3-2 z) z+\xi  \left(4 z^3-12 z^2+z+4\right)\right)\right) \delta^{ij}_\perp (\epsilon^{AV i}_\perp \Delta^j_\perp )}{Q \left(m^2_{AV} \left(-4 z^2+4
   z+1\right)+4 Q^2 z\right) \left(m^2_{AV} \left(2 x z+\xi  \left(4 z^2-2 z-1\right)\right)+2 Q^2 z (x-\xi )\right)} \\
%%%
& {\cal \widetilde{H}}^{TT}_{t2} = -\frac{4 m^2_{AV} \left(4 z^3-4 z^2-z+1\right)  \left(m^2_{AV}+Q^2\right) (\xi +x) \delta^{ij}_\perp (\epsilon^{AV i}_\perp \epsilon^{\gamma* j}_\perp)}{\left(m^2_{AV} \left(4 z^2-4 z-1\right)-4 Q^2 z\right) \left(m^2_{AV} \left(2 x z+\xi 
   \left(4 z^2-2 z-1\right)\right)+2 Q^2 z (x-\xi )\right)}
\end{align}
%%%%%%%%%%%%

%%%%%%%%%%%%%%%%%%%%%%%%%%%%%%%%%%%%%%%%%%%%%%%%%%%%%  
\subsection*{Diagram \texorpdfstring{$\boldsymbol{(e)}$}{(e)}}
%%%%%%%%%%%%%%%%%%%%%%%%%%%%%%%%%%%%%%%%%%%%%%%%%%%%% 
For diagram~($e$), we find that, for the $F$-type GTMDs, the case of a longitudinally
polarized virtual photon and a longitudinally polarized meson yields a vanishing
contribution at both twist-2 and twist-3.
%%%%%%%%%%%%
\begin{align}
{\cal H}^{LL}_{t2}=0, \quad {\cal H}^{LL}_{t3, k_\perp} =0, \quad {\cal H}^{LL}_{t3, \Delta_\perp} =0.
\end{align}
%%%%%%%%%%%%
We next present the remaining results:
%%%%%%%%%%%
\begin{align}
{\cal H}^{TL}_{t3,\Delta_\perp} &= -\frac{4 i m_{AV} (2 z-1) \left(m^2_{AV}\left(x (2 z-1)+\xi  \left(8
   z^3-8 z^2-4 z+1\right)\right)+Q^2 (2 z-1) (x+\xi  (2 z-1))\right) \epsilon^{ij}_\perp (\epsilon^{\gamma* i}_\perp \Delta_\perp^j) }{\left(m^2_{AV} \left(4 z^2-4 z-1\right)+4 Q^2 (z-1)\right)
   \left(m^2_{AV} \left(2 x (z-1)+\xi  \left(4 z^2-6 z+1\right)\right)+2 Q^2 (z-1) (\xi +x)\right)} \\
%%%
{\cal H}^{LT}_{t3,\Delta_\perp} & = \frac{4 i m^2_{AV} (2 z-1)  \left(m^2_{AV} \left(2 z^2-z-1\right)
   (x+\xi  (2 z-1))+Q^2 \left(x \left(2 z^2-z-1\right)-\xi  \left(4 z^3-8 z^2+z-1\right)\right)\right) \epsilon^{ij}_\perp (\epsilon^{AV i}_\perp \Delta_\perp^j) }{Q \left(m^2_{AV} \left(4 z^2-4
   z-1\right)+4 Q^2 (z-1)\right) \left(m^2_{AV} \left(2 x (z-1)+\xi  \left(4 z^2-6 z+1\right)\right)+2 Q^2 (z-1) (\xi +x)\right)} \\
%%%
{\cal H}^{TT}_{t2} & = \frac{4 i m^2_{AV} z (2 z - 3) (2 z - 1)\left(m^2_{AV}+Q^2\right) (x-\xi ) \epsilon^{ij}_\perp (\epsilon^{AV i}_\perp \epsilon^{\gamma* j}_\perp) }{\left(m^2_{AV} \left(4 z^2-4 z-1\right)+4 Q^2 (z-1)\right) \left(m^2_{AV} \left(2 x
   (z-1)+\xi  \left(4 z^2-6 z+1\right)\right)+2 Q^2 (z-1) (\xi +x)\right)}
\end{align}
%%%%%%%%%%%

We now turn to the results for the G-type GTMDs. The results that are non-zero are:
%%%%%%%%%%%%
\begin{align}
& {\cal \widetilde{H}}^{LL}_{t2} =  \frac{2 m_{AV} (2 z-1) \left(m^2_{AV}+Q^2\right) \left(m^2_{AV} \left(2 x+\xi  \left(8 z^3-12 z^2+2 z-1\right)\right)+2 Q^2 \left(2
   z^2-3 z+1\right) (\xi +x)\right)}{Q \left(m^2_{AV} \left(4 z^2-4 z-1\right)+4 Q^2 (z-1)\right) \left(m^2_{AV}  \left(2 x (z-1)+\xi 
   \left(4 z^2-6 z+1\right)\right)+2 Q^2 (z-1) (\xi +x)\right)} \\
%%%
& {\cal \widetilde{H}}^{TL}_{t3,\Delta_\perp} =  \frac{4 m_{AV} (2 z-1) \left(m^2_{AV} \left(x (1-2 z)^2+\xi  \left(8
   z^3-12 z^2+2 z-1\right)\right)+Q^2 (2 z-1) (\xi +x (2 z-1))\right) \delta^{ij}_\perp (\epsilon^{\gamma* i}_\perp \Delta^j_\perp )}{\left(m^2_{AV} \left(4 z^2-4 z-1\right)+4 Q^2 (z-1)\right)
   \left(m^2_{AV} \left(2 x (z-1)+\xi  \left(4 z^2-6 z+1\right)\right)+2 Q^2 (z-1) (\xi +x)\right)} \nonumber \\
%%%
& {\cal \widetilde{H}}^{LT}_{t3,\Delta_\perp} =\nonumber \\
& \frac{4 m^2_{AV} (2 z-1)  \left(m^2_{AV} \left(x \left(-2
   z^2+z+1\right)+\xi  z \left(4 z^2-8 z+3\right)\right)+Q^2 \left(x \left(-2 z^2+z+1\right)+\xi  \left(4 z^3-11 z+3\right)\right)\right) \delta^{ij}_\perp (\epsilon^{AV i}_\perp \Delta^j_\perp )}{Q
   \left(m^2_{AV} \left(4 z^2-4 z-1\right)+4 Q^2 (z-1)\right) \left(m^2_{AV} \left(2 x (z-1)+\xi  \left(4 z^2-6 z+1\right)\right)+2 Q^2
   (z-1) (\xi +x)\right)} \\
%%%
& {\cal \widetilde{H}}^{TT}_{t2} = -\frac{4 m^2_{AV} z (-3 + 2 z) (-1 + 2 z)\left(m^2_{AV}+Q^2\right) (x-\xi ) \delta^{ij}_\perp (\epsilon^{AV i}_\perp \epsilon^{\gamma* j}_\perp) }{\left(m^2_{AV} \left(4 z^2-4 z-1\right)+4 Q^2 (z-1)\right) \left(m^2_{AV} \left(2 x
   (z-1)+\xi  \left(4 z^2-6 z+1\right)\right)+2 Q^2 (z-1) (\xi +x)\right)}
\end{align}
%%%%%%%%%%%%

%%%%%%%%%%%%%%%%%%%%%%%%%%%%%%%%%%%%%%%%%%%%%%%%%%%%%  
\subsection*{Diagram \texorpdfstring{$\boldsymbol{(f)}$}{(f)}}
%%%%%%%%%%%%%%%%%%%%%%%%%%%%%%%%%%%%%%%%%%%%%%%%%%%%% 
For diagram~($f$), we find that, for the $F$-type GTMDs, the case of a longitudinally
polarized virtual photon and a longitudinally polarized meson yields a vanishing
contribution at both twist-2 and twist-3.
%%%%%%%%%%%%
\begin{align}
{\cal H}^{LL}_{t2}=0, \quad {\cal H}^{LL}_{t3, k_\perp} =0, \quad {\cal H}^{LL}_{t3, \Delta_\perp} =0.
\end{align}
%%%%%%%%%%%%
We next present the remaining results:
%%%%%%%%%%%%
\begin{align}
{\cal H}^{TL}_{t3,\Delta_\perp} & = -\frac{4 i m_{AV} (2 z-1)\left(m^2_{AV} \left(x (2 z-1)+\xi  \left(8
   z^3-16 z^2+4 z+3\right)\right)+Q^2 (2 z-1) (\xi +x-2 \xi  z)\right) \epsilon^{ij}_\perp (\epsilon^{\gamma* i}_\perp \Delta_\perp^j) }{\left(m^2_{AV} \left(4 z^2-4 z-1\right)-4 Q^2 z\right)
   \left(m^2_{AV} \left(\xi +2 x z-4 \xi  z^2+2 \xi  z\right)+2 Q^2 z (\xi +x)\right)} \\
%%%
{\cal H}^{LT}_{t3,\Delta_\perp} & = -\frac{4 i m^2_{AV} (2 z-1) \left(Q^2 \left(x (3-2 z) z+\xi  \left(-4
   z^3+4 z^2+3 z-4\right)\right)-m^2_{AV} z (2 z-3) (\xi +x-2 \xi  z)\right) \epsilon^{ij}_\perp (\epsilon^{AV i}_\perp \Delta_\perp^j) }{Q \left(m^2_{AV} \left(-4 z^2+4 z+1\right)+4 Q^2 z\right)
   \left(m^2_{AV} \left(\xi +2 x z-4 \xi  z^2+2 \xi  z\right)+2 Q^2 z (\xi +x)\right)} \\
%%%
{\cal H}^{TT}_{t2} & = -\frac{4 i m^2_{AV} (-1 - z + 2  z^2) (2 z - 1) \left(m^2_{AV}+Q^2\right) (x-\xi ) \epsilon^{ij}_\perp (\epsilon^{AV i}_\perp \epsilon^{\gamma* j}_\perp) }{\left(m^2_{AV} \left(4 z^2-4 z-1\right)-4 Q^2 z\right) \left(m^2_{AV} \left(\xi +2 x
   z-4 \xi  z^2+2 \xi  z\right)+2 Q^2 z (\xi +x)\right)}
\end{align}
%%%%%%%%%%%%

We now turn to the results for the G-type GTMDs. The results that are non-zero are:
%%%%%%%%%%%%
\begin{align}
& {\cal \widetilde{H}}^{LL}_{t2} =  \frac{2 m_{AV} (2 z-1) \left(\text{mv}^2+Q^2\right) \left(m^2_{AV} \left(2 x-\xi  \left(8 z^3-12 z^2+2 z+3\right)\right)+2 Q^2 z (2
   z-1) (\xi +x)\right)}{Q \left(m^2_{AV} \left(-4 z^2+4 z+1\right)+4 Q^2 z\right) \left(m^2_{AV} \left(\xi +2 x z-4 \xi  z^2+2 \xi 
   z\right)+2 Q^2 z (\xi +x)\right)} 
   \end{align}
   \begin{align}
& {\cal \widetilde{H}}^{TL}_{t3,\Delta_\perp} = \frac{4 m_{AV} (2 z-1) \left(m^2_{AV} \left(\xi  \left(8 z^3-12
   z^2+2 z+3\right)-x (1-2 z)^2\right)-Q^2 \left(\xi +x (1-2 z)^2-2 \xi  z\right)\right) \delta^{ij}_\perp (\epsilon^{\gamma* i}_\perp \Delta^j_\perp ) }{\left(m^2_{AV} \left(4 z^2-4 z-1\right)-4 Q^2
   z\right) \left(m^2_{AV} \left(\xi +2 x z-4 \xi  z^2+2 \xi  z\right)+2 Q^2 z (\xi +x)\right)} \\
%%%
& {\cal \widetilde{H}}^{LT}_{t3,\Delta_\perp} \nonumber \\
&=  -\frac{4 m^2_{AV} (2 z-1)\left(m^2_{AV} \left(x z (2 z-3)+\xi 
   \left(4 z^3-4 z^2-z+1\right)\right)+Q^2 \left(x z (2 z-3)+\xi  \left(4 z^3-12 z^2+z+4\right)\right)\right) \delta^{ij}_\perp (\epsilon^{AV i}_\perp \Delta^j_\perp ) }{Q \left(m^2_{AV} \left(-4 z^2+4
   z+1\right)+4 Q^2 z\right) \left(m^2_{AV} \left(\xi +2 x z-4 \xi  z^2+2 \xi  z\right)+2 Q^2 z (\xi +x)\right)} \\
%%%
& {\cal \widetilde{H}}^{TT}_{t2} = \frac{4 m^2_{AV} (-1 + z) (-1 + 2 z) (1 + 2 z) \left(m^2_{AV}+Q^2\right) (x-\xi ) \delta^{ij}_\perp (\epsilon^{AV i}_\perp \epsilon^{\gamma* j}_\perp) }{\left(m^2_{AV} \left(4 z^2-4 z-1\right)-4 Q^2 z\right) \left(m^2_{AV} \left(\xi +2 x
   z-4 \xi  z^2+2 \xi  z\right)+2 Q^2 z (\xi +x)\right)}
\end{align}
%%%%%%%%%%%%
We close this section by noting two points. First, as discussed in the main text and evident from the results above, all contributions vanish at the symmetric point $z = 1/2$. Second, concerning the Ward identity, we emphasize that it is satisfied either at $z = 1/2$ or when the meson mass is neglected. If both $z \neq 1/2$ and a finite meson mass are retained, the Ward identity is no longer satisfied. For completeness, we have presented here the corresponding results for the axial-vector meson obtained when both effects are retained (i.e.\ $z \neq 1/2$ and $m_{AV} \neq 0$); these expressions do not satisfy the Ward identity. We will continue to investigate in future work how to construct results that preserve the Ward identity while simultaneously retaining $z \neq 1/2$ and a finite meson mass.

%%%%%%%%%%%%%%%%%%%%%%%%%%%%%%%%%%%%%%%%%%%%%%%%%%%%% 
\section{GPD limit of vector meson results: comparison with the literature}
\label{appa}
%%%%%%%%%%%%%%%%%%%%%%%%%%%%%%%%%%%%%%%%%%%%%%%%%%%%% 
%%%%%%%%%%%%%
In this appendix, we begin by explicitly displaying the GPD parametrization equations. Using our hard-kernel results, we then present the amplitudes at twist-2 accuracy written directly in terms of GPDs. This formulation enables a transparent comparison with the existing literature, where GPD-based analyses have been studied extensively. We restrict our discussion to vector meson production only. Throughout the presentation below, we follow the notation introduced in the previous appendix.

The unpolarized gluon GPDs are defined as~\cite{Meissner:2007rx}
%%%%%%%%%%%%%%%%%%%%%%
\begin{align}
\int d^2k_\perp \, W^g
=
\frac{1}{2P^+}\,
\bar{u}(p', \lambda')\,
\bigg[
\gamma^+\, H^g
+
\frac{i\,\sigma^{+\mu}\Delta_\mu}{2M}\, E^g
\bigg]\,
u(p, \lambda) \, ,
\end{align}
%%%%%%%%%%%%%%%%%%%%%%
while the polarized (helicity-dependent) gluon GPDs are given by
%%%%%%%%%%%%%%%%%%%%%%
\begin{align}
\int d^2k_\perp \, \widetilde{W}^g
=
\frac{1}{2P^+}\,
\bar{u}(p', \lambda')\,
\bigg[
\gamma^+ \gamma_5\, \widetilde{H}^g
+
\frac{\Delta^+ \gamma_5}{2M}\, \widetilde{E}^g
\bigg]\,
u(p, \lambda) \, .
\end{align}
%%%%%%%%%%%%%%%%%%%%%%

At twist-2 accuracy, contributions arise only from longitudinally polarized photon–meson transitions and from transversely polarized photon–meson transitions. A direct inspection shows that, at this level, only the $H^g$ and $E^g$ GPDs contribute. The GPDs
$\widetilde{H}^g$ and $\widetilde{E}^g$ enter first at twist-3, where the hard
kernel is proportional to $\Delta_\perp$ and contributes to crossed
photon-meson polarization transitions, as shown in the main text. The corresponding results are:
%%%%%%%%%%%%
\begin{align}
    {\cal M}^{LL}_{t2} &= - C \int^1_{-1} \dfrac{dx}{(x-\xi+ i \varepsilon) (x+\xi- i \varepsilon)} \dfrac{Q }{m_V} \sqrt{1-\xi^2} \bigg ( H^g - \dfrac{\xi^2}{1-\xi^2} E^g \bigg ) , \\
    %%%%%
    {\cal \widetilde{M}}^{LL}_{t2} &= - C \int^1_{-1} \dfrac{dx}{(x-\xi+ i \varepsilon) (x+\xi- i \varepsilon)} \dfrac{Q}{m_V}\dfrac{\sqrt{-t'}}{2 M} \lambda \, E^g , \\
    %%%%
    {\cal M}^{TT}_{t2} &= C \int^1_{-1} \dfrac{dx}{(x-\xi+ i \varepsilon) (x+\xi- i \varepsilon)} (\delta^{ij}_\perp \epsilon^{V i}_\perp \epsilon^{\gamma*j}_\perp)  \sqrt{1-\xi^2} \bigg ( H^g - \dfrac{\xi^2}{1-\xi^2} E^g \bigg ) , \\
    %%%%%
    {\cal \widetilde{M}}^{TT}_{t2} &=  C \int^1_{-1} \dfrac{dx}{(x-\xi+ i \varepsilon) (x+\xi- i \varepsilon)}\dfrac{\sqrt{-t'}}{2 M} \lambda \, E^g , \\
\end{align}
%%%%%%%%%%%%
where 
%%%%%%%%%%%%
\begin{equation}
\sqrt{-t'} \equiv \sqrt{-(t - t_0)} = \frac{|\vec{\Delta}_\perp|}{\sqrt{1 - \xi^2}},
\end{equation}
%%%%%%%%%%%%
where $t_0$ denotes the minimal kinematically allowed value of $t$, and $C$ is given in Eq.~(\ref{e:c_vector}). As in the main text, we fix the transverse momentum transfer to $
\Delta_\perp = (|\vec{\Delta}_\perp|,\, 0)$.
Our results are in agreement with Ref.~\cite{Koempel:2011rc}, up to an overall sign difference in the terms involving the GPD $E$, which can be traced to different sign conventions associated with the choice of light-cone momentum direction for the proton.

%%%%%%%%%%%%%%%%%%%%%%%%%%%%%%%%%%%%%%%%%%%%%%%%%%%%%   
\bibliography{refs}

\end{document}